\documentclass[aip, reprint, twocolumn, amsmath,amsfonts,amssymb]{revtex4-1}
\usepackage{graphicx}
\usepackage{amsmath,amsfonts,amssymb}
\usepackage[colorlinks=true,linkcolor=blue]{hyperref}
\usepackage{color}
\usepackage{bm}
\usepackage{threeparttable}
\usepackage[mathscr]{eucal}
\usepackage{picture}

\usepackage{color}
\usepackage{bm}
\definecolor{grey}{rgb}{0.7,0.7,0.7}
\newcommand{\black}{\color{black}{}}

\usepackage{fancyvrb}
\usepackage{ulem}

\usepackage[T1]{fontenc}
\usepackage{braket}

\begin{document}
\title{Adaptive construction of shallower quantum circuits with quantum spin projection for fermionic systems}
\author{Takashi Tsuchimochi}
\email{tsuchimochi@gmail.com}
\affiliation{Graduate School of System Informatics, Kobe University, 1-1 Rokkodai-cho, Nada-ku, Kobe, Hyogo 657-8501 Japan}
\affiliation{Japan Science and Technology Agency (JST), Precursory Research for Embryonic Science and Technology (PRESTO), 4-1-8 Honcho Kawaguchi, Saitama 332-0012 Japan}
\author{Masaki Taii}
\affiliation{Graduate School of Science, Technology, and Innovation, Kobe University, 1-1 Rokkodai-cho, Nada-ku, Kobe, Hyogo 657-8501 Japan}
\author{Taisei Nishimaki}
\affiliation{Graduate School of System Informatics, Kobe University, 1-1 Rokkodai-cho, Nada-ku, Kobe, Hyogo 657-8501 Japan}
\author{Seiichiro L. Ten-no}
\affiliation{Graduate School of Science, Technology, and Innovation, Kobe University, 1-1 Rokkodai-cho, Nada-ku, Kobe, Hyogo 657-8501 Japan}
\affiliation{Graduate School of System Informatics, Kobe University, 1-1 Rokkodai-cho, Nada-ku, Kobe, Hyogo 657-8501 Japan}

\begin{abstract}
Quantum computing is a promising approach to harnessing strong correlation in molecular systems; however, current devices only allow for hybrid quantum-classical algorithms with a shallow circuit depth, such as the variational quantum eigensolver (VQE). In this study, we report the importance of the Hamiltonian symmetry in constructing VQE circuits adaptively. This treatment often violates symmetry, thereby deteriorating the convergence of fidelity to the exact solution, and ultimately resulting in deeper circuits. We demonstrate that symmetry-projection can provide a simple yet effective solution to this problem, by keeping the quantum state in the correct symmetry space, to reduce the overall gate operations. The scheme also reveals the significance of preserving symmetry in computing molecular properties, as demonstrated in our illustrative calculations.
\end{abstract}
\maketitle

\section{Introduction}
Recent advances in quantum technology have attracted widespread attention in various disciplines. Electronic structure theory is one of the fields expected to potentially benefit from the development of quantum computers because strong correlation can be handled by mapping the wave function directly onto entangled qubits without an exponential increase in computational cost. However, current noisy intermediate-scale quantum (NISQ) devices cannot implement error correction owing to the limited quantum resources; hence, they suffer from errors triggered by noise and short coherent times,\cite{Preskill18, Arute19, McArdle20} which limit the application of general quantum algorithms such as quantum phase estimation. Therefore, several hybrid quantum-classical algorithms have been proposed to address the technical challenges emerging in NISQ hardware. The variational quantum eigensolver (VQE) is one such algorithm,\cite{Peruzzo14, McClean16} in which a quantum state is propagated by a parametrized short quantum circuit that can be optimized in classical post processing, to variationally lower the energy expectation value and obtain the Hamiltonian ground state. 

In quantum chemistry, the most conventional ansatz for VQE circuit is the unitary coupled-cluster (UCC),\cite{Kutzelnigg82, Kutzelnigg84, Bartlett89, Romero19, Sokolov20} where fermionic excitation operators with respective to the Fermi vacuum (e.g. Hartree-Fock) are anti-hermitized and exponentiated to form unitary gates. Given the remarkable success of standard CC in a wide range of classical applications, UCC with single and double substitutions (UCCSD) is expected to be a feasible starting point for chemical Hamiltonians. Nevertheless, UCCSD is manifestly incapable of describing strong correlations,\cite{Lee19} which is the primary target of quantum computing. Extending the excitation manifold to include general orbitals, generalized UCCSD captures higher excitation effects in a compact manner, and is thus known to often provide accurate results for strongly correlated systems.\cite{Wecker15, Lee19} However, these UCC-based methods are inhibited by a limited representability because of the fixed structure of unitary gates; in addition, they also require several parameters and deep circuits, which makes them difficult for NISQ devices to handle.

Ansatz-free algorithms are more flexible and can control the accuracy and circuit depth in a highly black-box manner.\cite{Grimsley19, Ryabinkin20, Tang21, Gomes21, Yordanov21} A prominent algorithm in this context is ADAPT-VQE proposed by Grimsley et al.,\cite{Grimsley19} which adaptively adds a set of short unitary gates to the quantum circuit successively. Each unitary is chosen from an operator pool comprising generalized-UCCSD excitation operators, based on the VQE energy gradient. Later, Tang et al. proposed qubit-ADAPT\cite{Tang21}, which employs Pauli rotations instead of a UCC-like operator to drastically reduce the gate depth. Recently, Yordanov et al. demonstrated that ``qubit-excitations,'' each of which is a linear combination of Pauli strings in qubit-ADAPT, can be implemented with only a handful of CNOT gates.\cite{Yordanov20} Based on this result, they employed qubit-excitations to form an operator pool in what they called qubit-excitation based (QEB) ADAPT.\cite{Yordanov21} However, these ADAPT-based methods break important symmetries of chemical Hamiltonians, of which the most significant is spin-symmetry.

It is well known in the quantum chemistry community that breaking spin-symmetry triggers several undesired consequences, which include inaccurate energies and properties, a slow convergence to the exact state, and a difficulty in interpreting obtained results.\cite{Handy85, Andrews91, Tsuchimochi10B} ADAPT-based methods result in spin-symmetry breakages because of the use of fermionic or qubit excitations, for which spin-adaptation is difficult when exponentiated. This is more the rule than the exception. In fact, in strongly correlated systems, spin-symmetry is completely broken because ADAPT typically requires large rotation parameters to represent the multi-configuration character. As will be demonstrated, this causes the slow convergence of ADAPT algorithms and inaccurate results with truncated operators.

To address this issue, this study employs a spin-projection circuit,\cite{Tsuchimochi20} which restores the total spin quantum number to restrict ADAPT and explore the operator candidate in the correct symmetry space. We demonstrate that our algorithm can generally reduce the circuit depth, although it increases the measurement cost. In particular, we show that the spin-projected ADAPT outperforms UCCSD as a better tradeoff between accuracy and circuit depth, which is not always guaranteed in ADAPT. To investigate the importance of spin-symmetry in molecular property calculations, we further derive the first-order energy derivative in the presence of the spin-projection operator. 

The remainder of this paper is organized as follows. In Sections \ref{sec:FermionicADAPT} and \ref{sec:qubitADAPT}, we review the original algorithms for ADAPT-VQE, and describe how they break symmetries in Section \ref{sec:spin-contamination}. Section~\ref{sec:SP-ADAPT} presents detailed discussions on the application of spin-projection to each ADAPT scheme, including an explanation on how the simulated qubits can be tapered off in the presence of spin-projection. Subsequently, we derive the energy derivative for ADAPT in Section~\ref{sec:properties}, especially focusing on the importance of orbital response. In Section~\ref{sec:symmetry-breaking}, we demonstrate that the convergence in ADAPT becomes increasingly slower for strongly correlated systems, and in Section~\ref{sec:C}, we present our benchmark results on the spin-projected ADAPT, comparing the accuracy and gate efficiency between different methods. Section~\ref{sec:pools} demonstrates how the number of operators in a pool can be reduced in QEB ADAPT. The calculations for dipole moment and geometry are explained in Section~\ref{sec:property_results}. Finally, we conclude this work in Section~\ref{sec:conclusions}.

\section{Theory}\label{sec:Theory}
\subsection{Fermionic ADAPT-VQE}\label{sec:FermionicADAPT}

First, we review the ADAPT-VQE algorithm originally proposed by Grimsley et al.\cite{Grimsley19} This algorithm is based on the observation that the exact full configuration interaction (FCI) state can be reached from the Hartree-Fock (HF) state using the exponential form,
\begin{align}
	|\Psi^{\rm FCI}\rangle = \prod_k e^{t_k\hat \tau(k)} |\Phi^{\rm HF}\rangle \label{eq:FCI}
\end{align}
where $\hat\tau(k)$ is the $k$th instance of arbitrarily ordered operators comprising the anti-symmetrized single and double excitations,
\begin{subequations}
\begin{align}
&	\hat \tau_q^p = a^\dag_pa_q - a^\dag_q a_p\\
&	\hat \tau_{rs}^{pq} = a^\dag_p a^\dag_q a_r a_s - a^\dag_s a^\dag_r a_q a_p
\end{align} 	
\label{eq:tau}
\end{subequations}
Here, we have adopted $p>q,r>s$ to denote general spin-orbitals. Note that the same anti-symmetrized excitation operators can appear repeatedly in Eq. \ref{eq:FCI}, but with different variational parameters $t_k$. Although Eq.~(\ref{eq:FCI}) includes only singles and doubles, the effect of higher excitations such as triples and quadruples can be conveniently considered by their exponential products.

Regardless of its exactness, Eq.~\ref{eq:FCI} is extremely cumbersome and long because it requires a large number of $k$; hence, its realization is prohibitively difficult on quantum computers. On the one hand, from a circuit complexity perspective, such as the number of CNOT gates, one can employ only a limited number of unitaries $e^{t_k \hat \tau(k)}$ on NISQ computers. On the other hand, it is expected that such a truncated ansatz with a reasonable number $n$ of unitaries can still outperform fixed ans\"atze such as UCCSD, especially for strongly correlated systems, with  appropriately chosen $\hat \tau(k) \; (k=1,\cdots,n)$ and variationally optimized $t_k$. Therefore, the objective of ADAPT-VQE is to dynamically create an ansatz that approaches FCI, using a maximally compact sequence of $n$ unitary operators, which are successively determined based on the energy gradient. To do so, an operator pool ${\mathscr P}$ is first defined, which comprises $M$ operators that are used to construct the trial state:
\begin{align}
			{\mathscr P} = \{\hat{\tau}^{p_\alpha}_{q_\alpha}+\hat{ \tau}^{p_\beta}_{q_\beta}, \hat{\tau}^{p_\alpha q_\alpha}_{r_\alpha s_\alpha}+\hat{\tau}^{{p}_\beta {q}_\beta}_{{r}_\beta s_\beta} ,\hat{\tau}^{p_\alpha q_\beta}_{r_\alpha s_\beta}+\hat{\tau}^{p_\beta q_\alpha}_{r_\beta s_\alpha}\} \label{eq:Pool}
\end{align}
where we explicitly consider the spins $\alpha$ and $\beta$ and each linear combination of operators is labeled by $m=1,\cdots, M$. Note that each combination of doubles does not comprise the spin-complement form, which we would address later in this paper. 

A trial state of ADAPT-VQE at the $n$th cycle is given by
\begin{align}
	\Ket{\psi_n}=e^{\theta_n \hat{\cal A}_n}e^{\theta_{n-1} \hat{\cal A}_{n-1}} \cdots e^{\theta_1 \hat{\cal A}_1}\Ket{\psi^{\rm HF}} \label{eq:ADAPT}
\end{align}
where $\hat {\cal A}_k \in {\mathscr P}$, and the parameters $\{\theta_k\}$ are variationally optimized to minimize the energy expectation value. To select the most contributing operator from the ${\mathscr P}$ for $n+1$th cycle, we create trial states with each $\hat A_m \in {\mathscr P} \; (m=1,\cdots, M)$ as
	\begin{align}
	|\psi_{n+1, m}\rangle = e^{\theta_{n+1} \hat A_m} |\psi_{n}\rangle
	\end{align}
and calculate the energy gradient for each $|\psi_{n+1,m}\rangle$ around $\theta_{n+1} = 0$,
\begin{align}
	R_{m}^{(n)} = \left. \frac{\partial \langle \psi_n | e^{-\theta \hat A_m} \hat H e^{\theta \hat A_m}| \psi_n \rangle}{\partial \theta}\right|_{\theta = 0} = \langle \psi_n | \left[\hat H, \hat A_m\right] | \psi_n \rangle\label{eq:Rm}
\end{align}
which can be evaluated with up to three-body reduced density matrix (3RDM) because of the commutator property.
The operator with the largest gradient $R_m^{(n)}$ is selected as $\hat {\cal A}_{n+1}$, with which the parameters $\{\theta_k: k=1,\cdots,n+1\}$ in $|\psi_{n+1}\rangle$ are all optimized by the standard VQE.

The ADAPT algorithm presented above is converged when the norm of Eq. (\ref{eq:Rm}) is smaller than the threshold $\epsilon$. It is worth noting that such a condition is closely related to the $k$-particle generalized Brillouin theorem.\cite{Kutzelnigg79} Kutzelnigg demonstrated that an exact FCI wave function satisfies $\langle \Psi_{\rm FCI}| \left[\hat H, \hat \tau_\mu\right] |\Psi_{\rm FCI}\rangle = 0$ for all excitation ranks. The convergence in ADAPT-VQE with $\epsilon=0$ only corresponds to the 1- and 2-particle generalized Brillouin conditions, and therefore, ADAPT-VQE is not guaranteed to converge to FCI. However, in practice, it approaches the FCI accuracy as $||{\bf R}||$ decreases.\cite{Claudino20}

\subsection{Qubit-based ADAPT-VQE}\label{sec:qubitADAPT}
\begin{figure*}
	\includegraphics[width = 55 em]{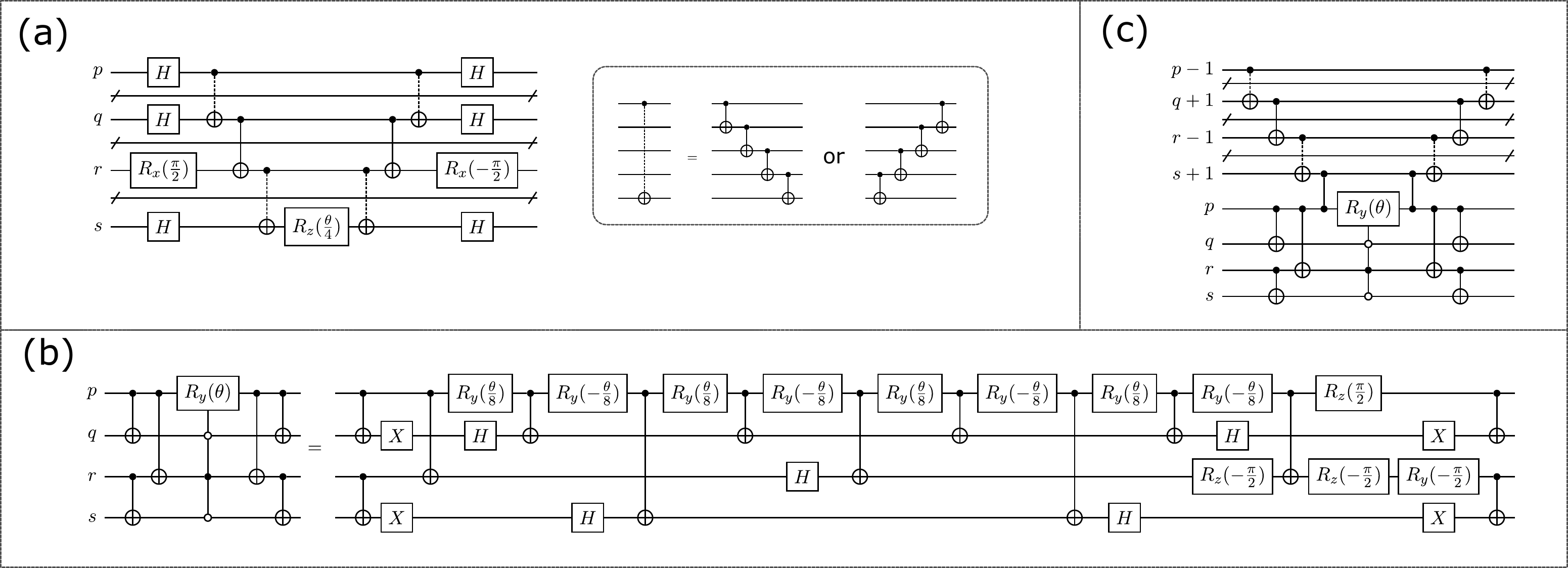}
	\caption{Quantum circuits for (a) $e^{-\frac{i\theta}{8} X_pX_qY_rX_s}$, where the CNOT with a dashed line is a shorthand for the CNOT ladder to convey the parities (refer to the inset), (b) $e^{\theta \hat {\tilde \tau}_{rs}^{pq}}$, and (c) $e^{\theta \hat {\tau}_{rs}^{pq}}$ using the controlled-Ry gate.}\label{fig:circ}
\end{figure*}

The aforementioned Fermionic ADAPT is based on the Jordan-Wigner transformation for mapping fermionic excitation operators Eqs.~(\ref{eq:tau}) to Pauli operators. This introduces the chain of $Z$ operations because of the anti-symmetric character of fermions, e.g.,
\begin{align}
	\hat \tau_{rs}^{pq}&= \frac{i}{8} \big(Y_p X_q X_r X_s + X_p Y_q X_r X_s - X_p X_q Y_r X_s \nonumber\\
	& - X_p X_q X_r Y_s - X_p Y_q Y_r Y_s - Y_p X_q Y_r Y_s \nonumber\\
	&+ Y_p Y_q X_r Y_s + Y_p Y_q Y_r X_s\big) \bigotimes_{t=q+1}^{p-1} Z_{t} \bigotimes_{u=s+1}^{r-1} Z_{u}\label{eq:JW}
\end{align} 
where we assume $p>q>r>s$, but similar relations can be obtained for other cases. 
The exponential of $\hat \tau_{rs}^{pq}$ is usually implemented by decomposing the eight Pauli strings presented in Eq.~\ref{eq:JW}, each for which requires the ladders of CNOT gates to take into account the parities for $t = q+1, \cdots, p-1$ and $u = s+1, \cdots, r-1$; refer to Fig.~\ref{fig:circ} as an example. Therefore, such a na\"ive implementation of $e^{\theta \hat \tau_{rs}^{pq}}$ necessitates $8(p-q+r-s+2)\ge 48$ CNOTs. However, several studies \cite{Gomes20, Xia20, Zhang21} have demonstrated that even if all the $Z$ operators from the Pauli string in Eq.~(\ref{eq:JW}) are discarded to define the qubit excitation
\begin{align}
\hat{\tilde \tau}_{rs}^{pq} &= \frac{i}{8} \big(Y_p X_q X_r X_s + X_p Y_q X_r X_s - X_p X_q Y_r X_s \nonumber\\
	& - X_p X_q X_r Y_s - X_p Y_q Y_r Y_s - Y_p X_q Y_r Y_s \nonumber\\
	&+ Y_p Y_q X_r Y_s + Y_p Y_q Y_r X_s\big),\label{eq:tau_tilde}
\end{align}
the accuracy obtained by VQE will be similar to that with the full $\hat \tau_{rs}^{pq}$, although it eliminates a considerable number of CNOT gates.
Tang et al. attempted to further decompose $\hat{\tilde \tau}_{rs}^{pq}$ and construct a reduced operator pool for what they called qubit-ADAPT-VQE,\label{Tang21}, noting that each Pauli string has a similar action. Using circuits depicted in Fig.~\ref{fig:circ}(a), each operator $e^{\theta_m\hat A_m}$ requires only six CNOT gates (similarly, two CNOT gates for single excitation variants, $Y_p X_q$). However, as will be discussed, this treatment significantly breaks several symmetries in the Hamiltonian and slows down the algorithm’s convergence for chemical applications. Moreover, because it violates the number symmetry, it cannot be applied to ionized states or half-filled Hubbard models. 

Recently, Yordanov et al. demonstrated that the role of $e^{\theta \hat {\tilde \tau}_{rs}^{pq}}$ is to rotate the bit strings between $|0_p0_q1_r1_s\rangle$ and $|1_p1_q0_r0_s\rangle$ by $\theta$, while everything else is unaffected, and proposed the application of a controlled-Ry gate, as illustrated in Fig.~\ref{fig:circ}(b), up to the phase.\cite{Yordanov20} This implementation requires only 13 CNOT gates, although the connectivity between $pqrs$ is assumed. They have also established that a standard double excitation $e^{\theta \tau_{rs}^{pq}}$ can be similarly implemented (Fig.~\ref{fig:circ}(c)), where the number of CNOT gates is reduced to $2(p-q+r-s)+9$. Using Eq.~(\ref{eq:tau_tilde}) and Fig.~\ref{fig:circ}(b), they proposed qubit-excitation-based (QEB) ADAPT, which was shown to perform significantly better than qubit-ADAPT of Ref.~[\onlinecite{Tang21}] in terms of CNOT gate counts and number of parameters.\cite{Yordanov21} 

\subsection{Symmetry-breaking in ADAPT-VQE} \label{sec:spin-contamination}
In the original proposal by Grimsley et al., it appears to be convenient to choose $\hat{\tau}^{p_\alpha q_\alpha}_{r_\alpha s_\alpha}+\hat{\tau}^{{p}_\beta {q}_\beta}_{{r}_\beta s_\beta}$ and $\hat{\tau}^{p_\alpha q_\beta}_{r_\alpha s_\beta}+\hat{\tau}^{p_\beta q_\alpha}_
{r_\beta s_\alpha}$ as the elements of the operator pool for fermionic ADAPT, because operators in each linear combination are symmetric in terms of spin. However, they are neither generally commutative ($\hat \tau^{p_\alpha q_\beta}_{r_\alpha s_\beta}$ and $\hat{\tau}^{p_\beta q_\alpha}_
{r_\beta s_\alpha}$ do not commute if $p=q$ or $r=s$) nor spin-free. This results in two limitations. First, because the operator selection is based on Eq.~(\ref{eq:Rm}) while the VQE part (implicitly) introduces the Trotter approximation, the gradients between these two steps can be inconsistent, which may cause an occasional numerical instability . In our experience, the same operator can be chosen between consecutive ADAPT cycles, and the convergence to FCI can never be achieved in some cases. Second, the application of a spin-dependent operator basis sacrifices the conservation of spin $s$. Although a more recent study has employed the spin-free doubles generator, that is, the sum of these two combinations,\cite{Tang21} it necessitates an undesired Trotter decomposition, as demonstrated in our previous study.\cite{Tsuchimochi20} Using such spin-incomplete operators provokes a serious issue widely known in electronic structure as spin-contamination. This is especially problematic in strongly correlated systems where the initial HF state significantly deviates from the FCI state and therefore large amplitudes $\theta_k$ are essential. Certainly, a lost spin-symmetry would be eventually recovered by the ADAPT algorithm, provided the target spin state is the ground state and the aforementioned instability is circumvented. However, large spin-symmetry breaking is inevitable in the early steps of the ADAPT procedure, which often leads to unphysical gradients and a poor convergence, as will be demonstrated below. The qubit-ADAPT algorithm\cite{Tang21} is also expected to suffer from the same problem. Furthermore, the candidate operators in qubit-ADAPT break other two important symmetries, $\hat S_z$ and $\hat N$, and the problem becomes even more severe. This fact can be easily confirmed by transforming each of the strings in Eq.\ref{eq:tau_tilde} back to their corresponding fermionic representations; for example, for the case where $p=6, q=3, r=1, s=0$, we have
\begin{align}
\frac{i}{8} Y_6 X_3 X_1 X_0=& \frac{1}{8} \big( a_6^\dag a_3^\dag a_1 a_0+ a_6^\dag a_3^\dag a_1^\dag a_0^\dag \nonumber\\&-2a_6^\dag a_4^\dag a_1^\dag a_0^\dag a_4 a_3+\cdots- h.c.\big).
\end{align}

Hence, it is important to maintain a correct symmetry throughout the ADAPT algorithm, as much as possible, to prevent any unnecessary diversion toward FCI. Accordingly, this study investigates the importance of symmetry-conserving sequence of operators, with a special focus on spin-symmetry.
 One may employ the constraint approach where $\lambda \left(\hat O - o\right)^2$ is added to the energy functional as a penalty to enforce the target state to be an (approximate) eigenfunction of the symmetry operator $\hat O$ with the eigenvalue $o$. Here, we take a different approach and restore the broken symmetry in each step of ADAPT by adopting appropriate post treatments, that is, spin-projection.

\subsection{Spin-projected ADAPT}\label{sec:SP-ADAPT}
In electronic structure theory, spin-projection is a vital tool,\cite{Lowdin55A, Scuseria11, Jimenez12, Tsuchimochi16A, Tsuchimochi18} and has recently exhibited a significant potential in quantum computing in the context of VQE.\cite{Tsuchimochi20} It restores the spin of the broken-symmetry quantum state $|\psi\rangle$, where $\hat S^2 |\psi\rangle \ne s(s+1)|\psi\rangle$, by applying a spin-projection operator $\hat P$, such that 
\begin{align}
	\hat S^2 \hat P|\psi\rangle &= s(s+1)\hat P|\psi\rangle,\\
	\hat S_z \hat P|\psi\rangle &= m_s\hat P|\psi\rangle,
\end{align} 
where $m_s = (N_{\alpha} - N_\beta)/2$.
Following our previous work, we express $\hat P$ as a linear combination of unitary operators,
\begin{align}
	\hat P \approx \sum_g^{N_g} w_g \hat U_g,\label{eq:P}
\end{align}
where $\hat U_g = e^{-i \alpha_g \hat S_z} e^{-i \beta_g \hat S_y} e^{-i \gamma_g \hat S_z}$ and $w_g$ are weights defined by the Wigner D-matrix. In VQE, we are interested in the expectation value of energy, 
\begin{align}
	E = \frac{\langle \psi|\hat P^\dag \hat H \hat P|\psi\rangle}{\langle \psi|\hat P^\dag\hat P|\psi\rangle} \approx \frac{\sum_g w_g\langle \psi|\hat H \hat U_g|\psi\rangle}{\sum_g w_g\langle \psi|\hat U_g|\psi\rangle}\label{eq:E}
\end{align}
where we have used $\hat P^2 = \hat P = \hat P^\dag$ and $\left[\hat H, \hat P\right] = 0$. Note that, using transfer operators $|s;m\rangle \langle s;k|$, which generate a spin $s$ state with $m_s = m$ from $m_s = k$, a more general projector $\sum_{m_s'} c_{m_s'} |s;m_s\rangle \langle s;m_s'|$ introduces more variational freedom, provided $|\psi\rangle$ is not an eigenstate of $\hat S_z$, by diagonalizing the Hamiltonian to determine the coefficients $c_k$\cite{Jimenez12}; however, for simplicity, we only consider $|s;m_s\rangle \langle s;m_s|$, which results in Eqs.~\ref{eq:P} and \ref{eq:E}.
When $|\psi\rangle$ is an eigenstate of $\hat S_z$, Eq.~\ref{eq:E} can be further simplified using the commutativity between $\hat H$ and $\hat S_z$, and one only needs to evaluate the Hamiltonian coupling $\langle \psi|\hat H e^{-i\beta_g \hat S_y}|\psi\rangle$ and the overlap $\langle \psi|e^{-i\beta_g \hat S_y}|\psi\rangle$. These expectation values are estimated, for instance, by the Hadamard test with one ancilla qubit, with a constant number of CNOTs ($4n_{\rm qubit}$). With the Gauss-Legendre quadrature, we take $N_g=2$ in the applications below; hence, the overall measurements will be doubled, while the circuit length is almost unchanged with or without $\hat P$.

The basic idea here is to introduce $\hat P$ to ADAPT-VQE and constrain the operator search in the correct spin space for a better convergence. Accordingly, we simply define the spin-projected (SP) ADAPT by applying $\hat P$ to Eq.~\ref{eq:ADAPT}.
\begin{align}
	|\psi^{\rm SP}_n\rangle = \hat P \; e^{\theta_n \hat{\cal A}_n}e^{\theta_{n-1} \hat{\cal A}_{n-1}} \cdots e^{\theta_1 \hat{\cal A}_1}\Ket{\psi^{\rm HF}}. \label{eq:SP-fADAPT}
\end{align}
The energy gradient evaluated to find the $n+1$th operator is then similar to Eq.~(\ref{eq:Rm}) as follows:
\begin{align}
R_m^{(n)} &= \left.\frac{\partial }{\partial \theta} \frac{\langle \psi_n| e^{-\theta \hat {\cal A}_m} \hat H \hat P \; e^{\theta \hat {\cal A}_m }|\psi_n\rangle}{\langle \psi_n| e^{-\theta \hat {\cal A}_m} \hat P \; e^{\theta \hat {\cal A}_m }|\psi_n\rangle} \right|_{\theta = 0}\nonumber\\
&= \frac{\langle \psi_n| \left[(\hat H - E) \hat P, \hat A_m\right] |\psi_n\rangle}{\langle \psi_n| \hat P |\psi_n\rangle} \label{eq:RmSP}
\end{align}
The commutator involves higher rank operators owing to the presence of $\hat P$, and $R_m^{(n)}$ requires four-body transition RDM between $|\psi_n\rangle$ and $\hat U_g|\psi_n\rangle$. However, in practice, they can be handled by the (Fermionic-)shift rule without explicitly computing RDMs\cite{Schuld19, Kottmann21, Izmaylov21} consistently, both in the operator selection and VQE steps.

The operator candidates used in the original fermionic ADAPT algorithm are pairwise in terms of spin, because if only either of the spin-dependent excitations is employed in ADAPT-VQE, the other would become almost indispensable to restore the spin symmetry. However, when $\hat P$ is introduced, such a pairwise treatment is no longer necessary. For example, consider the single excitation operators $\hat \tau_{q_\alpha}^{p_\alpha}$ and $\hat \tau_{q_\beta}^{p_\beta}$. Although each operator excites (and de-excites) an electron from the $q$-th orbital to the $p$-th orbital, the role of $\hat \tau_{q_\beta}^{p_\beta}$ is simply to complement the lost spin in $\hat \tau_{q_\alpha}^{p_\alpha}$ and {\it vice versa}, which can be handled by $\hat P$. Based on this consideration, it is advisable to employ completely spin-dependent excitation operators as a pool for Fermionic ADAPT ansatz,
\begin{align}
				{\mathscr P}^{\rm spin} = \{\hat{\tau}^{p_\alpha}_{q_\alpha}, \hat{ \tau}^{p_\beta}_{q_\beta}, \hat{\tau}^{p_\alpha q_\alpha}_{r_\alpha s_\alpha}, \hat{\tau}^{{p}_\beta {q}_\beta}_{{r}_\beta s_\beta} ,\hat{\tau}^{p_\alpha q_\beta}_{r_\alpha s_\beta},\hat{\tau}^{p_\beta q_\alpha}_{r_\beta s_\alpha}\} \label{eq:SPPool}
\end{align} 
In the presence of $\hat P$, the role of $\hat{\tau}^{p_\alpha q_\beta}_{r_\alpha s_\beta}$ becomes less significant if its spin-flipped operator $\hat{\tau}^{p_\beta q_\alpha}_{r_\beta s_\alpha}$ is already treated in the ansatz. Therefore, some spin-flipped operators that primarily produce effects similar to $\hat P$ would not be chosen in the algorithm, which will potentially minimize the overall circuit depth. Furthermore, the exponential of each operator in the pool ${\mathscr P}^{\rm spin}$ can be treated without the Trotter decomposition; therefore, the energy derivatives estimated in the VQE part and Eq.~(\ref{eq:RmSP}) are equivalent for the last chosen operator. This eliminates the algorithmic problem discussed in Section \ref{sec:spin-contamination}.

Spin-projection is also applicable to qubit-based ADAPT. However, as we have discussed, qubit-ADAPT, which adopts the decomposed Pauli strings as candidate operators, violates the $\hat S_z$ symmetry, and therefore prevents us from using the simplified $e^{-i\beta_g\hat S_y}$ rotation for spin-projection. The full rotation operator $ e^{-i \alpha_g \hat S_z} e^{-i \beta_g \hat S_y} e^{-i \gamma_g \hat S_z}$ allows one to project the correct $m_s$ state but entails both $\alpha$ and $\gamma$ rotation grids, which results in a drastic increase in the number of measurements, by two orders of magnitudes. 
In contrast, in QEB-ADAPT, $\hat {\tilde \tau}_{rs}^{pq}$ preserves the $\hat S_z$ and $\hat N$ symmetries. Because the difference between $\hat {\tilde \tau}_{rs}^{pq}$ and $\hat \tau_{rs}^{pq}$ is simply that the former neglects the parities between qubits, we can express it as
\begin{align}
	\hat {\tilde \tau}_{rs}^{pq} = \hat \tau_{rs}^{pq} + \sum_t c_t \hat \tau_{rst}^{pqt} + \sum_{tu} c_{tu} \hat \tau_{rstu}^{pqtu} +\cdots \label{eq:tau_tilde_expand}
\end{align}
where $t,u$ are the qubit indices appearing as a $Z$ string in Eq.~(\ref{eq:tau}), that is, $t,u = [q+1,p-1], [s+1,r-1]$. As a concrete example, it can be verified that
\begin{align}
	\hat {\tilde\tau}_{10}^{63} = \hat {\tau}_{10}^{63} - 2 \hat \tau_{104}^{634} - 2 \hat \tau_{105}^{635} - 4 \hat \tau_{1054}^{6354}.
\end{align}
Hence, spin-projection can be easily combined with QEB-ADAPT, and we call such an algorithm SP-QEB.

Finally, let us briefly consider the tapering-off technique for spin-projection. As it is known, because chemical Hamiltonians possess number and point-group symmetries, one can identify unitaries that transform a Hamiltonian such that it has only Pauli operators that trivially act on certain qubits, which can thus be discarded.\cite{Bravyi17, Setia20} Such unitaries are identified using the $\mathbb{Z}_2$ symmetry; namely, the parities of $\alpha$ and $\beta$ electron numbers for the number symmetry,\cite{Bravyi17} and the sign changes of the underlying wave function by (Abelian) point-group symmetry operations.\cite{Setia20} However, the spin-rotation operator $e^{-i\beta\hat S_y}$ changes the number parity of $\alpha$ and $\beta$ electrons as follows: 
\begin{align}
	\hat S_y = \frac{1}{2i}\sum_{p}\left(a_{p_\alpha}^\dag a_{p_\beta} - a_{p_\beta}^\dag a_{p_\alpha}\right)
\end{align}
Therefore, the $\mathbb{Z}_2$-symmetry can be exploited only for the total number operator $\hat N = \hat N_\alpha +\hat N_\beta$, but not for $\hat S_z = \frac{1}{2}(\hat N_\alpha - \hat N_\beta)$  because $\left[\hat S_y, \hat S_z\right] \ne 0$, resulting in the reduction of only one qubit instead of two. This is a necessary cost for spin-projection to exert its advantages; note that the $\hat S^2$ symmetry is not categorized as a $\mathbb{
Z}_2$ symmetry and the tapering-off scheme is not applicable. From a different perspective, spin-projection can explore a ``hidden'' Hilbert space that is not accessible by standard UCC-like ans\"atze by deliberately breaking and restoring the number symmetry of each electron spin. Although one cannot use number parity to discard the other one qubit from the simulation, there is no such restriction for the point-group symmetry. In some of our calculations discussed below, we taper qubits to ease the computational cost. However, we will assume the number of CNOT gates $N_{\rm CNOT}$, which is central to measuring the circuit complexity in this study, remains the same. In addition, we will not discuss how gate operations would change by the transformation in the tapering-off algorithm.

\subsection{First-order molecular properties}\label{sec:properties}
In molecular systems, one is often interested not only in the total energy but also chemical properties. Time-independent molecular properties are defined as the total energy change with respect to a perturbation $x$ introduced to the Hamiltonian. To obtain properties from quantum computing, several studies have focused on evaluating analytical energy derivatives.\cite{OBrien19, Mitarai20} Although higher-order properties such as polarizability require higher-order derivatives, and therefore, are difficult to compute, first-order properties $dE[x]/ dx$ such as force and dipole moment can be easily obtained in VQE. This is because the quantum state is variational with respect to VQE parameters $\{\theta_k\}$ used in the quantum circuit, 
\begin{align}
	\frac{\partial E[{\bm\theta}]}{\partial \theta_k} = 0. \label{eq:dEdtheta}
\end{align}
However, in both ADAPT-VQE and SP-ADAPT-VQE, the canonical orbitals of HF commonly employed as a starting point are generally not optimal. The fully parametrized wave function is given by $ \hat P e^{\hat \kappa} |\psi_{\rm ADAPT}[{\bm\theta}]\rangle$ ($\hat P$ is discarded for standard ADAPT-VQE), where 
\begin{align}
\hat \kappa = \sum_{p>q} \kappa_{pq} \hat \tau_q^p.
\end{align}
Hence, the total energy
\begin{align}
	E[{\bm\theta}, {\bm\kappa}, x] = \frac{\langle \psi_{\rm ADAPT}[{\bm\theta}] | e^{-\hat \kappa}\hat H[x] \hat P e^{\hat \kappa} |\psi_{\rm ADAPT}[{\bm\theta}]\rangle}{\langle \psi_{\rm ADAPT}[{\bm\theta}] | e^{-\hat \kappa} \hat P e^{\hat \kappa} |\psi_{\rm ADAPT}[{\bm\theta}]\rangle}
\end{align}
is not usually fully stationary with respect to orbital change
\begin{align}
	\left.\frac{\partial E[{\bm\theta},{\bm\kappa},x]}{\partial \kappa_{pq}} \right|_{{\bm\kappa} = {\bf 0}} \ne 0\label{eq:dEdk}
\end{align}
unless it is fully converged to the FCI state. In other words, the 1-particle generalized Brillouin condition is not satisfied; note that the left side of Eq.~(\ref{eq:dEdk}) is equivalent to the gradient $R_{pq}$ defined in Eqs.~(\ref{eq:Rm}) and (\ref{eq:RmSP}) for standard ADAPT and SP-ADAPT ans\"atze, respectively. To address this issue, we can compute the changes in molecular orbitals induced by $x$, by solving the coupled-perturbed HF equation\cite{Gerratt68, Pople79} or the Z-vector equation,\cite{Handy84} similar to previous studies.\cite{OBrien19, Mitarai20} In this study, we adopt an alternative (but mathematically equivalent) formulation commonly used in quantum chemistry,\cite{Helgaker00, Helgaker12} and rederive the necessary equations to obtain first-order molecular properties in SP-ADAPT. 

Accordingly, we introduce the following Lagrangian,
\begin{align}
	{\cal L}[{\bm\theta},{\bm\kappa}, {\bf z}, x] = E[{\bm\theta},{\bm\kappa},x] + \sum_{p>q}z_{pq} F_{pq} \label{eq:L}
\end{align}
where $z_{pq}$ and $F_{pq}$ are Lagrange multipliers to be determined and off-diagonal elements of the canonical Fock matrix, respectively. We note that the canonical HF orbitals require $F_{pq} = 0$ for $p\ne q$ at each $x$, and therefore, ${\cal L}$ is stationary with respect to $z_{pq}$. Evidently, for any $x$, ${\cal L}[{\bm\theta},{\bm\kappa}, {\bf z}, x]$ reproduces the same value as $E[{\bm\theta},{\bm\kappa},x]$, and therefore, $d{\cal L}/dx \equiv dE/dx$; however, ${\cal L}$ has an added advantage in that it can be stationary with respect to all variational parameters: ${\bm \theta}$, ${\bm \kappa}$, and {\bf z} (note that $\partial{\cal L}/\partial \theta_k =0$ is automatically satisfied by VQE, see Eq.~(\ref{eq:dEdtheta})). The stationary condition for ${\bm\kappa}$ can be achieved by solving the linear equation to determine ${\bf z}$: 
\begin{align}
\frac{	\partial {\cal L}[{\bm\theta},{\bm\kappa}, {\bf z}, x]}{\partial \kappa_{rs}} = R_{rs} + \sum_{p>q} z_{pq} A_{pq,rs} = 0 \label{eq:R+zA}
\end{align}
where
\begin{align}
	A_{pq,rs} = \frac{\partial F_{pq}}{\partial\kappa_{rs}}
\end{align}
is the Fock derivative with respect to orbital change and can be easily computed by a classical computer. We present the explicit form of {\bf A} and the solution to Eq.~(\ref{eq:R+zA}) in Appendix. 

Using the chain rule,
\begin{align}
	\frac{dE}{dx} & = \frac{d{\cal L}}{dx} \nonumber\\ &= \frac{\partial {\cal L}}{\partial x} + \sum_{k} \left(\frac{\partial {\cal L}}{\partial \theta_k}\frac{d\theta_k}{dx}\right) + \sum_{p>q} \left(\frac{\partial {\cal L}}{\partial \kappa_{pq}}\frac{d\kappa_{pq}}{dx} + \frac{\partial {\cal L}}{\partial z_{pq}}\frac{dz_{pq}}{dx} \right) \nonumber\\
	&= \frac{\partial {\cal L}}{\partial x} 
\end{align}
where the last term  is straightforward to evaluate once ${\bf z}$ is available. It is noteworthy  that this approach is applicable to higher order derivatives in a mathematically clear way, although we will not go into the details.

The Lagrangian ${\cal L}$ can be expressed by the so-called relaxed density matrices ${\bf D}^{\rm relax}$, which incorporate the response correction ascribed to the orbital change by perturbation (Appendix). It can be easily shown that molecular properties as energy derivatives are computed using the relaxed density matrices, instead of the expectation value of the corresponding observable operator, that is, with the regular density matrices $D_{pq} =\langle \psi|a^\dag_p a_q|\psi\rangle$. 
The simple expectation value using the latter is generally less accurate because the orbital response is neglected. 
To extend the orbital response correction to SP-ADAPT, we can resort to the derivation in Ref.~[\onlinecite{Tsuchimochi17A}].
In Appendix, we summarize the equations and also review the evaluation of nuclear gradients in the case where HF orbitals are a function of nuclear coordinates.

Although this approach can be applied to any VQE method, in several chemistry-inspired ansatzes such as UCCSD, $R_{rs}$ values are zero or negligible because single excitations may be explicitly treated in the variational optimization. However, we note that the frozen-core (or frozen-virtual) approximation is frequently exercised both in classical and quantum computing, where the lower core (higher virtual) orbitals are considered inert and fixed to those of the reference state that is often HF. Because these frozen orbitals are not explicitly optimized within VQE, a response correction is required. This can be easily achieved by simply expanding the range of orbital space $p,q$ in Eq.~(\ref{eq:L}) to include the frozen orbitals. In such cases, matrix elements with frozen orbitals, denoted by $I, J$, are required. Although the expression for $A_{pq,rI}$ remains the same, the gradient $R_{rI}$ needs to be handled in classical computers because $I$ is not mapped to qubits (refer to Appendix for comprehensive derivations). 
Certainly, all these problems would be eliminated if orbitals are optimized,\cite{Takeshita20, Mizukami20}; however, orbital-optimized VQE itself requires these quantities and extra VQE simulations. 

As mentioned above, for both fermionic and SP-fermionic ADAPTs, we have access to $R_{rs}^{(n)}$ at each $n$th cycle by construction. Therefore, the energy derivative and first-order molecular properties are readily available by processing density matrices in a classical computer. Conversely, for the qubit-based ADAPT, one needs to explicitly compute $R_{rs}$ unless the quantum state is exact. Otherwise, the property would be computed as an expectation value and can suffer from an uncontrollable error. 

\section{Illustrative calculations}

\begin{table}
\caption{Summary of ADAPT protocols.}\label{tb:pool}
    \begin{tabular}{ccccccc}
        \hline\hline
        Method & Pool& Symmetry &  \\
         \hline
        Fermionic& Eq.~(\ref{eq:Pool}) & $\hat S_z, \hat N$\\
        Spin-dependent Fermionic& Eq.~(\ref{eq:SPPool}) &  $\hat S_z, \hat N$\\
        Qubit& $\{X_pY_q, X_pX_qX_rY_s\}$ & None\\
        QEB & $\{{\hat {\tilde \tau}_q^p}, {\hat {\tilde \tau}_{rs}^{pq}}\}$ &  $\hat S_z, \hat N$\\
        \hline\hline
        \end{tabular}
\end{table}

\subsection{Computational details}
\black
Before discussing our numerical results, we describe the computational details. 
All simulations were conducted using Quantum Unified Kernel for Emulation ({\sc Quket}), developed by us, without the effect of noise.\cite{Quket} {\sc Quket} compiles several open-source libraries to generate a Hamiltonian mapped to a qubit basis and perform quantum simulations with parallel computing. Specifically, it employes {\sc PySCF}\cite{pyscf} to generate molecular orbitals and integrals, and {\sc OpenFermion}\cite{openfermion} to perform the Jordan-Wigner transformation. To simulate quantum circuits, we utilize {\sc Qulacs}.\cite{qulacs} The energy minimization in VQE uses the Broyden-Fletcher-Goldfarb-Shannon (BFGS) algorithm implemented in Scipy. The geometry optimization was performed by interfacing {\sc Quket} with {\sc pyberny}.\cite{pyberny}

For Fermionic-ADAPT, we adopt two different pools, Eqs.~(\ref{eq:Pool}) and (\ref{eq:SPPool}), and the controlled-Ry circuit to perform $e^{\theta \hat\tau_{rs}^{pq}}$ and $e^{\theta \hat\tau_q^p}$ and estimate the number of CNOT gates $N_{\rm CNOT}$ in Fig.~\ref{fig:circ}(c). If the same operator is chosen between two consecutive ADAPT steps because of a numerical error or Trotter error, we choose the operator with the next largest gradient. We will not consider the effect of noise in this work, but simply evaluate the expected performance.

We summarize the ADAPT protocols used in this study and the corresponding operator pools in Table~\ref{tb:pool}. For spin-projection, we always use the pool of either the spin-dependent Fermionic excitations or qubit excitations.

\begin{figure}
	\includegraphics[width = 27 em]{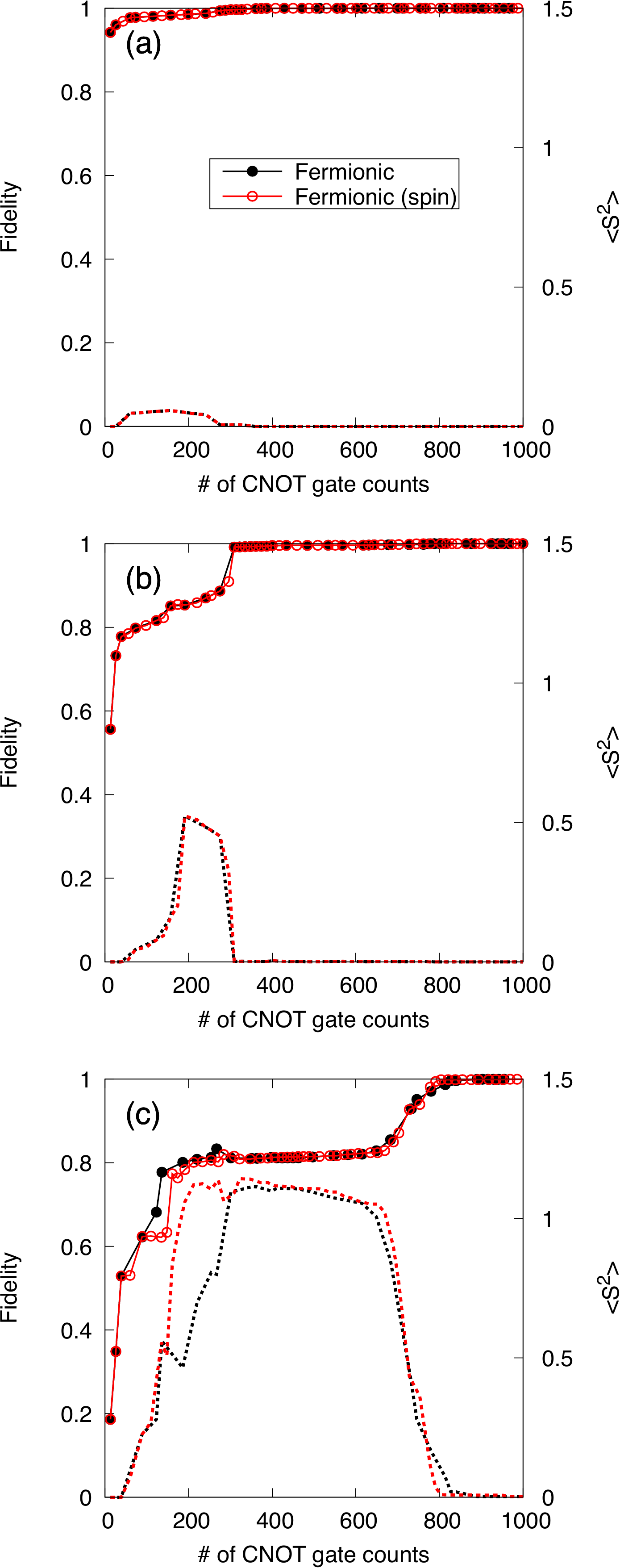}
	
	\caption{Fidelity (solid lines) and $\langle \hat S^2\rangle$ (dotted lines) of ADAPT-VQE, plotted against $N_{\rm CNOT}$ for N$_2$ with different bond lengths: (a) 1.098 {\AA}, (b) 1.8 {\AA}, and (c) 2.5 {\AA}. Circles indicate each ADAPT step.}\label{fig:fidelity}
\end{figure}
\subsection{Spontaneous symmetry breaking in ADAPT}\label{sec:symmetry-breaking}
First, we examine the spontaneous symmetry breaking and its consequence in ADAPT for the N$_2$ molecule with the STO-3G basis set. We have considered three different bond lengths, R$_{\rm N-N} = 1.098, 1.8$, and 2.5 {\AA} to represent weakly and strongly correlated cases. The solid lines in Fig.\ref{fig:fidelity} depict the fidelity of the Fermionic ADAPT-VQE state to the exact FCI state $|\langle \Psi_{\rm FCI}|\psi_{\rm ADAPT} \rangle|^2$ as a function of $N_{\rm CNOT}$, using either ${\mathscr P}$ or ${\mathscr P}^{\rm spin}$, labeled as Fermionic and Fermionic (spin), respectively. In all these cases, ADAPT-VQE approaches the exact state as more operators are added, as expected. The figure indicates that there is no significant difference in convergence between ${\mathscr P}$ and ${\mathscr P}^{\rm spin}$, although the former, the original algorithm proposed in Ref.[\onlinecite{Grimsley19}], is more advantageous in that it requires less VQE parameters, as indicated by the circles.
However, the convergence behavior is significantly different depending on the system. In other words, the stronger the correlation of a system, the slower the convergence of ADAPT-VQE to FCI. 

This behavior is attributed to the broken spin characterized by $\langle \hat S^2\rangle$, which is also illustrated in Fig. \ref{fig:fidelity} with the dotted lines. In the strongly correlated case of R$_{\rm N-N} = 2.5$ {\AA} (Fig.~\ref{fig:fidelity}(c)), the HF state exhibits a very small overlap with FCI, with a fidelity of 0.2, which quickly increases to 0.8 after the initial few ADAPT steps with approximately 200 CNOTs. However, simultaneously, $\langle \hat S^2\rangle$ (simply the degree of the spin-symmetry breaking) becomes larger than 1, and at this point, the fidelity convergence gets trapped in a plateau. 

Such large spin-symmetry breakings occur initially because, in the VQE step of ADAPT, large VQE parameters are essential for describing strong correlations (i.e., highly excited configurations from the HF reference). Consequently, the $\alpha$ and $\beta$ electrons are treated quite differently, especially in the absence of the Trotter decomposition.\cite{Tsuchimochi20} We should mention that the spin-symmetry is safely conserved if ``paired'' double excitations such as $\hat \tau_{q_\alpha q_\beta}^{p_\alpha p_\beta}$ are chosen. In fact, the first few operators selected in ADAPT are of this type, and $\langle\hat S^2\rangle$ remains as zero (see Figs.~\ref{fig:fidelity} and S1). However, paired double excitations do not have a sufficient ability to describe all types of correlation effects, and spin-unpaired excitations are more suitable in several cases.
With $\langle \hat S^2\rangle>1$, the ADAPT state is a mixture of several different spin states, each with a significant weight. Therefore, the derivative approach expressed in Eq.~(\ref{eq:Rm}) to determine the operator candidate is likely inappropriate, because such an operator search is performed in an incorrect symmetry space.  Of course, by further processing ADAPT with more operators and CNOT gates, the lost spin-symmetry starts to be restored ($\sim$700 CNOTs for Fig.~\ref{fig:fidelity}(c)). At this point, the fidelity also quickly increases to one. Similar but less pronounced results can be confirmed for the tests with shorter bond lengths. 

\begin{figure}
	\includegraphics[width = 25em]{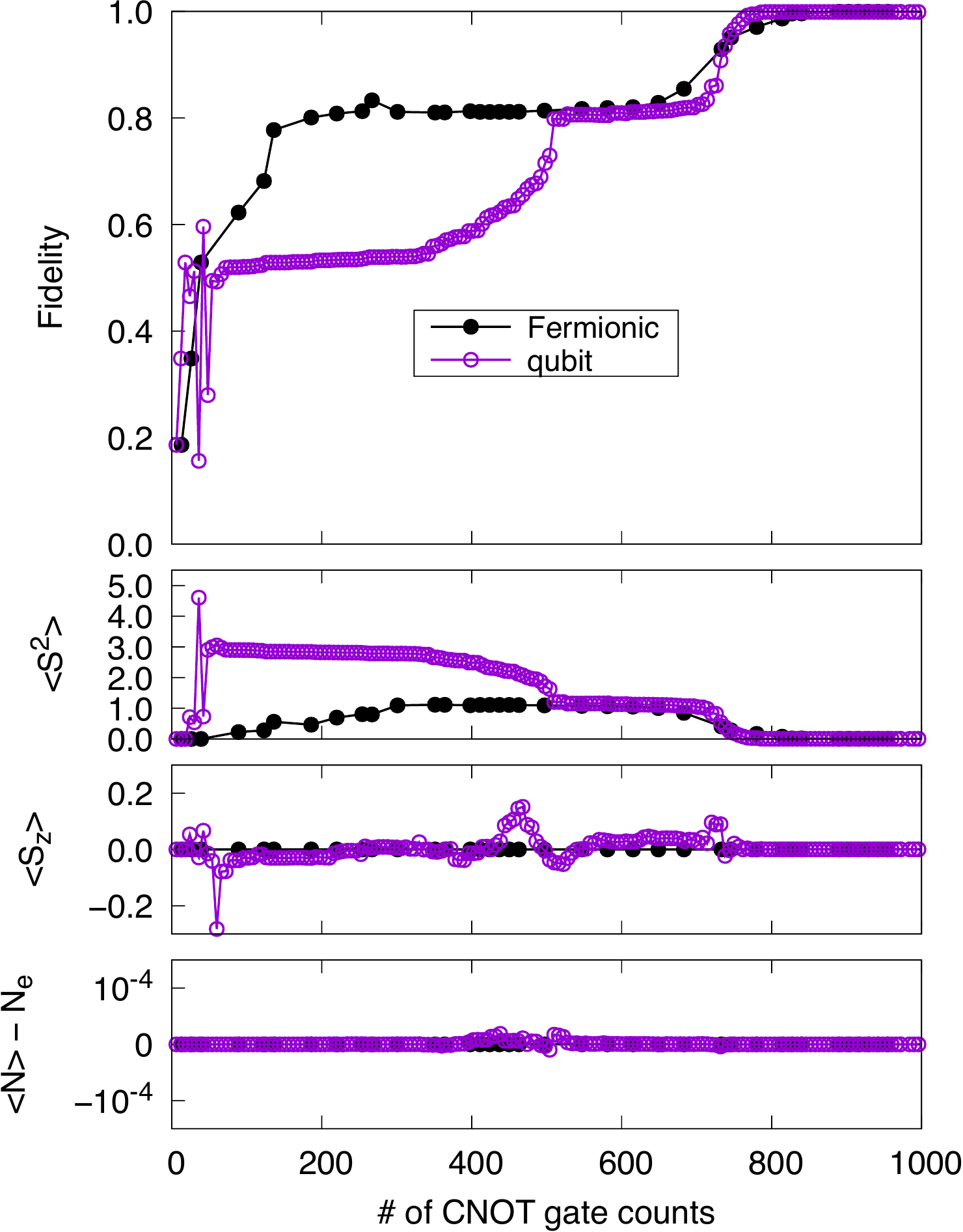}
	\caption{Comparison between fermionic-ADAPT and qubit-ADAPT for N$_2$ at R$_{\rm N-N}=$ 2.5 {\AA}. Circles indicate each ADAPT step.} \label{fig:N2_25}
\end{figure}

Although fermionic ADAPT follows the $\hat S_z$ and $\hat N$ symmetries, qubit-ADAPT breaks all, and as a result, exhibits a significant slow-down of convergence to FCI. This is illustrated by Fig.~\ref{fig:N2_25}, where we have plotted the fidelity and symmetry expectation values, $\langle \hat S^2\rangle, \langle \hat S_z\rangle,$ and $\langle \hat N\rangle$ using the problematic case of $N_2$ at R$_{\rm N-N} = 2.5$ {\AA}. We note that, in this particular case, both algorithms coincidentally result in similar CNOT gate counts to achieve the FCI state. However, the number of VQE parameters for Fermionic-ADAPT to achieve a fidelity of 0.99 is 30, while that for qubit-ADAPT is 128, implying that more measurements are required for qubit-ADAPT. Interestingly, the symmetry-breaking of qubit-ADAPT in $\hat S_z$ and $\hat N$ is rather moderate; however, the error in $\langle\hat S^2\rangle$ is substantially larger than that of Fermionic-ADAPT. Apparently, such a large error in $\langle\hat S^2\rangle$ severely affects the fidelity, and it is evident that the plateaus that emerge in the fidelity and $\langle\hat S^2\rangle$ plots correspond to each other.

\begin{figure*}
\includegraphics[width = 50em]{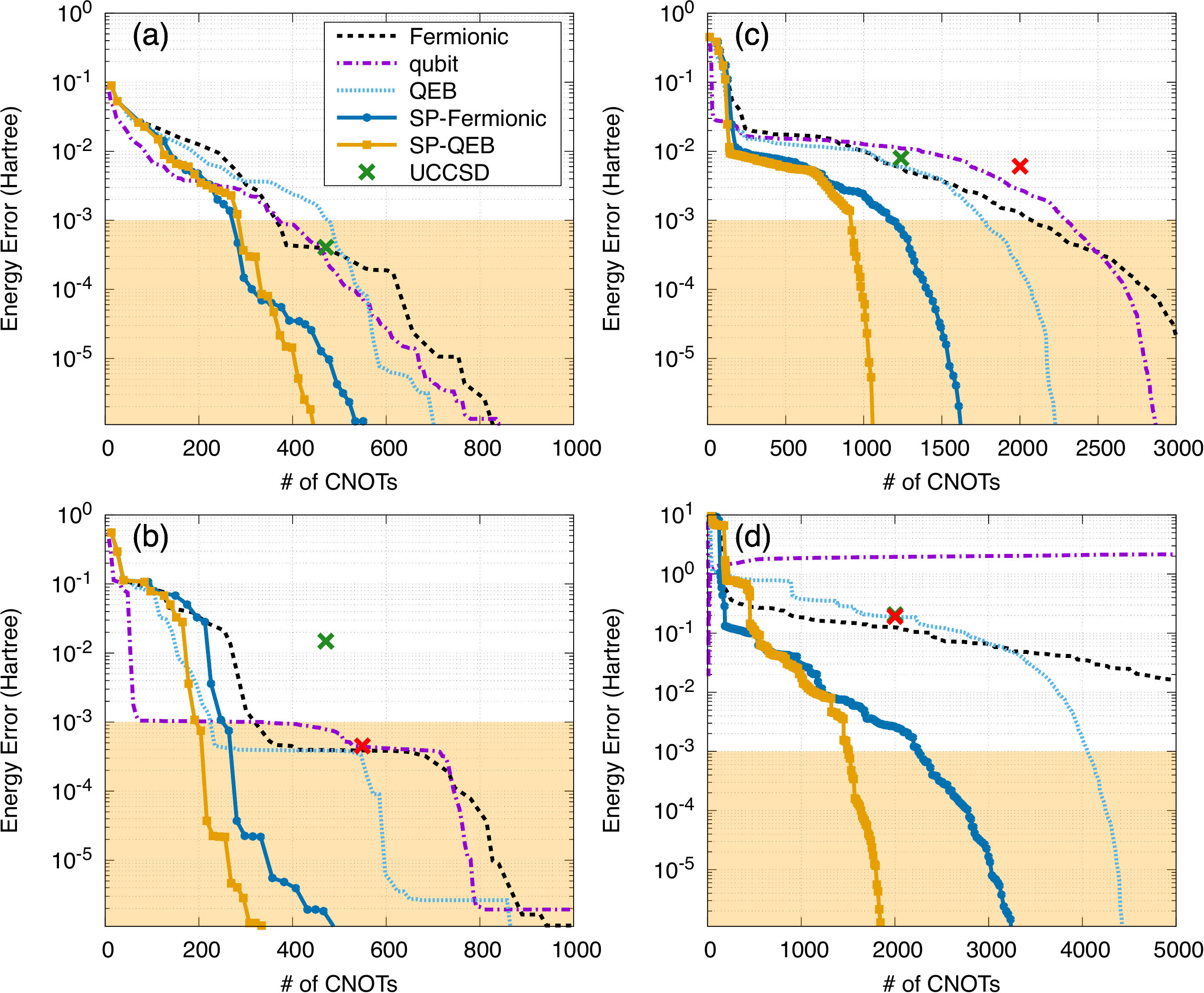}
\caption{Energy error from FCI (in Hartree) at each CNOT gate count. (a) N$_2$ at equilibrium geometry (R$_{\rm N-N} = 1.098$ {\AA}), (b) stretched N$_2$ (R$_{\rm N-N} = 2.5$ {\AA}), (c) stretched H$_6$ (R$_{\rm H-H}=2$ {\AA}), (d) half-filled 6-site Hubbard chain ($U=8$). The green and red plots indicate standard UCCSD and broken-symmetry UCCSD, respectively.}\label{fig:Error_vs_CNOT}
\end{figure*}

\begin{figure*}
	\includegraphics[width = 50em]{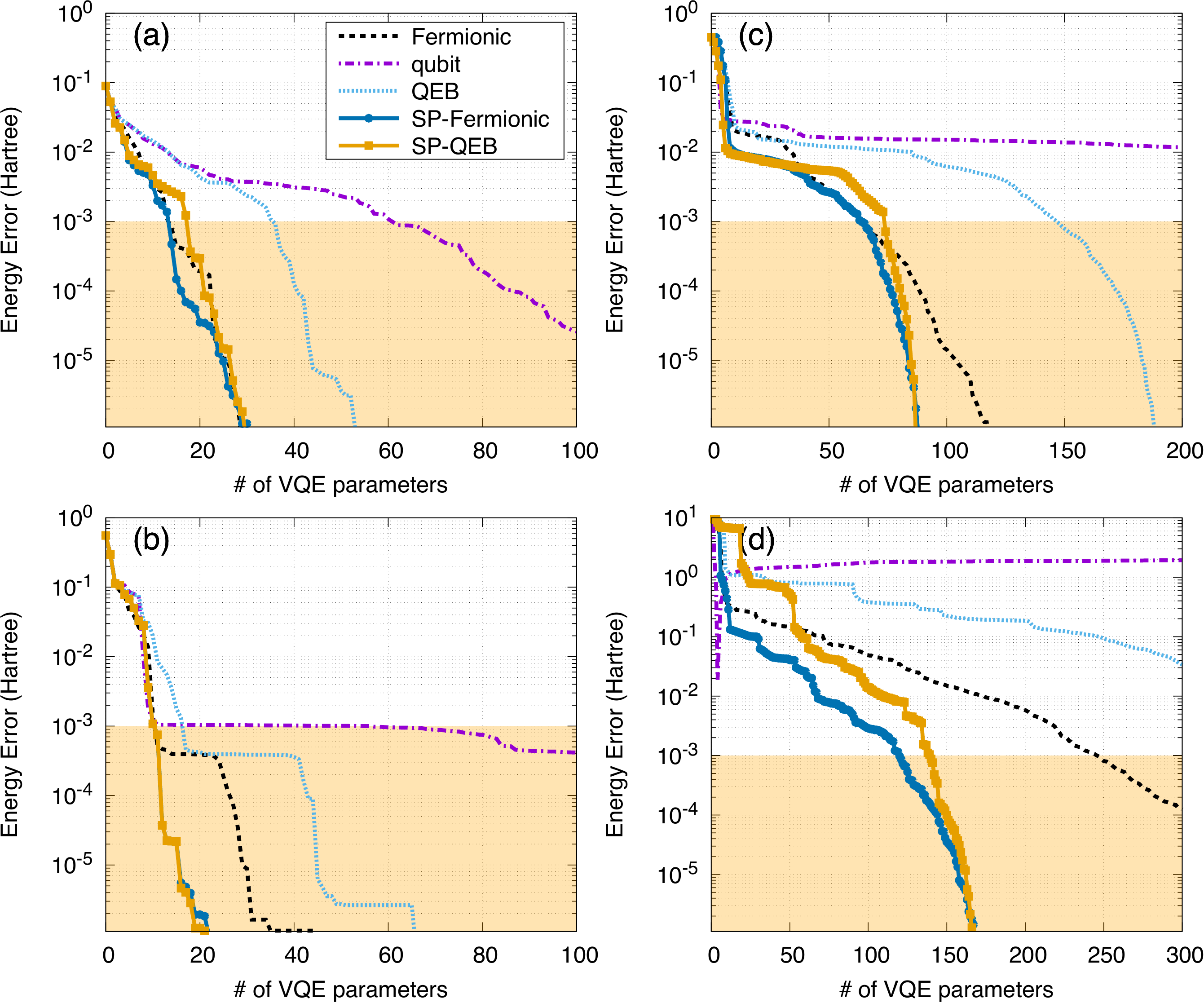}
	\caption{Same as Fig.~\ref{fig:Error_vs_CNOT} but as a function of the number of parameters.}\label{fig:Error_vs_param}
\end{figure*}
\begin{figure}
	\includegraphics[width = 25em]{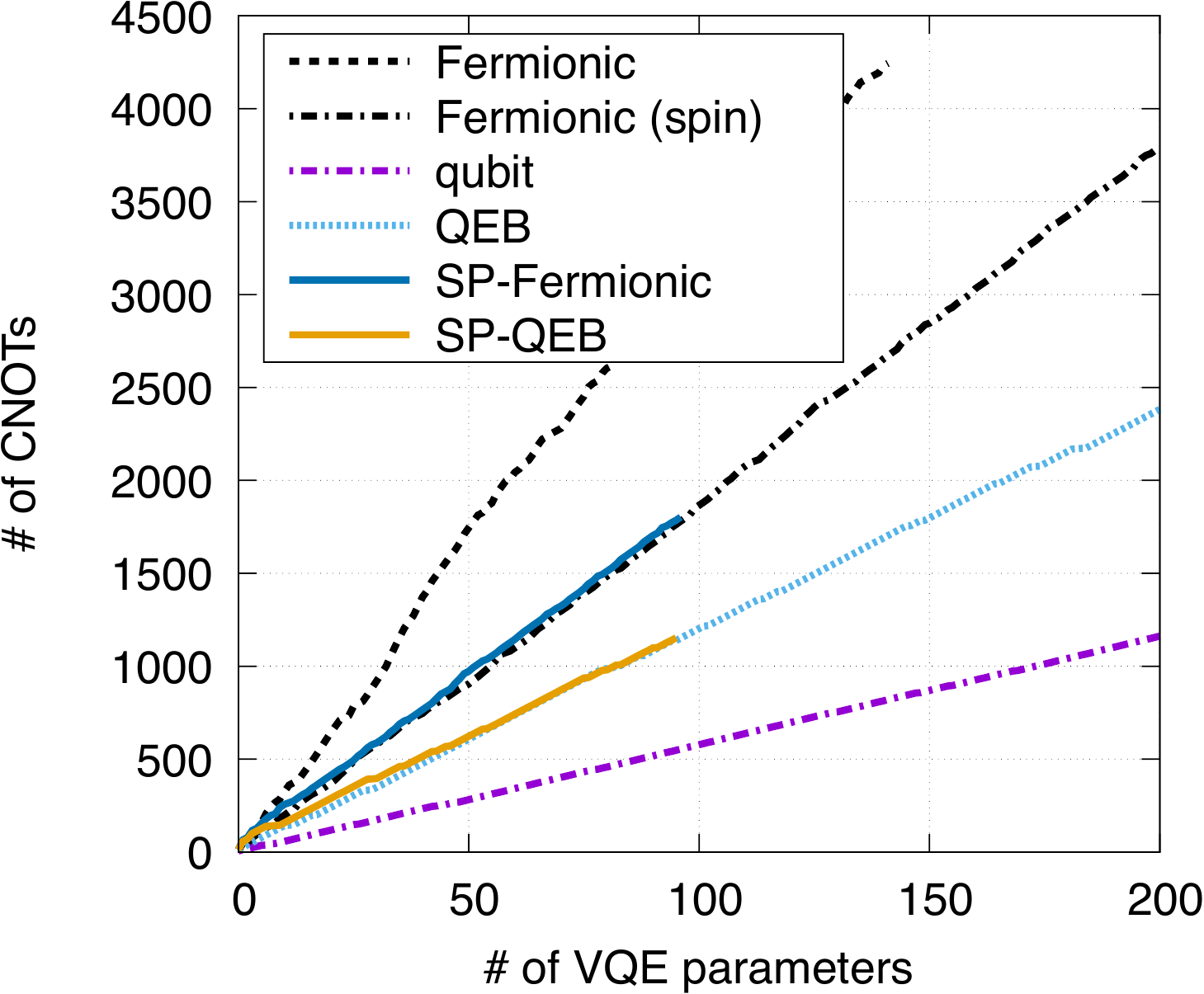}
	\caption{Increase in the number of CNOT gates with the number of parameters for H$_6$.
	}\label{fig:Nparam_vs_CNOT}
\end{figure}

\subsection{Accuracy and gate efficiency in ADAPT and SP-ADAPT}\label{sec:C}
As demonstrated above, symmetry-breaking is associated with strong correlation, and occurs spontaneously in the ADAPT algorithms, thereby slowing down the convergence. Therefore, one would naturally expect that maintaining the correct symmetry can mitigate the problem. In this section, we study the effect of spin-projection using several systems, which include the N$_2$ molecule (with R$_{N-N} = 1.098, 2.5$ {\AA}), linear chain of H$_6$ equally separated by 2 {\AA}, and half-filled one-dimensional periodic Fermi-Hubbard model with six sites and $U=8$. For the molecular systems, we have used the STO-3G basis set and six electrons in six orbitals. Therefore, all the models considered here are composed of 12 qubits. 

We choose the initial state to be the HF determinant for the molecular systems; however, for the Hubbard model, we intentionally set the charge localized state where the first three sites are doubly occupied and the remaining three sites are empty to break the spatial symmetry. 

Fig.~\ref{fig:Error_vs_CNOT} presents the energy error from FCI and the number of CNOT gates in the quantum circuit for several methods, where the shaded area in orange corresponds to chemical accuracy ($< 1$ mHartree error). We first consider the performance of qubit-ADAPT.\cite{Tang21} For the molecular systems (Figs.~\ref{fig:Error_vs_CNOT}(a-c)), no significant advantage is found over Fermionic ADAPT in terms of CNOT gate counts. This is because the current implementation of Fermionic ADAPT relies on the controlled-Ry circuit of Ref.~[\onlinecite{Yordanov20}], which can eliminate a large number of CNOT gates. In contrast, we find that qubit-ADAPT is disadvantageous in two aspects. First, the energy lowering in each VQE step is considerably small, as plotted in Fig.~\ref{fig:Error_vs_param}, and it requires several VQE parameters (measurements), which also makes it difficult to converge each VQE step because of a highly non-linear parameter space. Second, it breaks the electron number symmetry; the application of qubit-ADAPT to ionic cases is not straightforward, as illustrated by the example of the half-filled Hubbard model, where qubit-ADAPT falls into the {\it true} ground state of $\langle \hat N\rangle =4$ and $\langle \hat S^2\rangle = 2$. 

Now let us focus on the results obtained by spin-projection. Evidently, by eliminating the incorrect spin-symmetry components, SP-ADAPT methods (both for Fermionic and QEB) require much less CNOT gates to achieve the same accuracy as those without spin-projection. SP-Fermionic ADAPT is already effective for the weakly-correlated N$_2$ at equilibrium geometry, as illustrated in Fig.~\ref{fig:Error_vs_CNOT}(a). The CNOT gate reduction in this system with spin-projection (by a factor of $0.6\sim0.7$) is almost solely attributed to the fact that spin-complement excitations are handled by the superposition of spin-rotated states; in contrast, with Fermionic ADAPT, one needs to explicitly construct a quantum circuit to perform spin-complement excitations, except for paired doubles. This can be verified by plotting the energy error against the number of VQE parameters (Fig.~\ref{fig:Error_vs_param}). From Fig.~\ref{fig:Error_vs_param}(a), the number of VQE parameters required to achieve the same accuracy is quite similar between Fermionic and SP-Fermionic ADAPT algorithms, which implies the two algorithms yield similar quantum states for each ADAPT iteration. 

The efficacy of spin-projection becomes more distinct for strongly correlated systems. For the stretched N$_2$ and H$_6$ (Figs.~\ref{fig:Error_vs_CNOT}(b) and (c)), the number of CNOT gates required to reach the FCI ground state with SP-ADAPT is less than half of that with the corresponding broken-symmetry ADAPT. Having said that, spin-projection may not seem to be advantageous because it employs up to 200 CNOT gates. For instance, QEB-ADAPT exhibits a lower energy than SP-QEB-ADAPT in the stretched N$_2$ case, and the convergence of qubit-ADAPT is initially even faster. However, this is an artifact due to the unphysical spin contamination effect, and the quality of the quantum state is far from satisfactory. A large spin contamination also leads to incorrect properties and state assignment. Furthermore, as seen in Fig.~\ref{fig:N2_25}, again, it often causes the plateau. 

The number of required parameters in SP-ADAPT (Fig.~\ref{fig:Error_vs_param}) is less than half of standard ADAPT, as expected. As illustrated in Fig.~\ref{fig:Nparam_vs_CNOT}, the scaling of CNOT gate counts with the number of used parameters is essentially identical between ADAPT and SP-ADAPT when the same operator pool is used. Therefore, a lower number of parameters in the latter simply implies a more gate efficient circuit at each energy accuracy. SP-Fermionic ADAPT converges slightly faster than SP-QEB ADAPT; this may simply be because the former contains more operators in the pool. However, it is interesting to observe that both the SP-Fermionic and SP-QEB algorithms require almost the same number of parameters to represent the exact FCI. This suggests we can further reduce the number of operators to form a pool, which can reduce the measurements for derivative evaluations, while expecting an unchanged fine convergence behavior.

Finally, we also performed UCCSD to evaluate the advantages of ADAPT and SP-ADAPT. In our UCCSD implementation, only the gates with non-zero parameters are constructed to minimize the gate depth. Because the molecular systems we consider have a high spatial symmetry, most UCCSD excitations are symmetry-forbidden, thereby permitting gate-efficient circuits. With such a treatment, UCCSD is already as powerful as ADAPT for N$_2$ at equilibrium and H$_6$ (see the green plots in Figs.~\ref{fig:Error_vs_CNOT}(a) and (c)). For the stretched N$_2$, ADAPT appears to be significantly more efficient and accurate; however, this may well be attributed to the spin-symmetry breaking. For a fair comparison, we also present the broken-symmetry UCCSD\cite{Tsuchimochi20} in red, which provides a fully variational solution at the cost of large spin-contamination. In addition, note that the ``full'' UCCSD, which performs symmetry-forbidden gate operations, that is, even if the amplitudes are zero, can be prepared with 2001 CNOT gates for all cases. For the Hubbard model, UCCSD naturally requires all excitations, and thus, 2001 CNOT gates, as we have initiated the calculation with no spatial-symmetry; UCCSD’s performance is similar to QEB-ADAPT. In conclusion, although ADAPT is imminently more flexible than the fixed ansatz of UCCSD, it ``passes through'' a UCCSD-like state in the process of converging to the FCI state; this is quite logical because UCCSD is usually considered an accurate method that is both chemically and theoretically well established.

\begin{figure}
		\includegraphics[width = 25em]{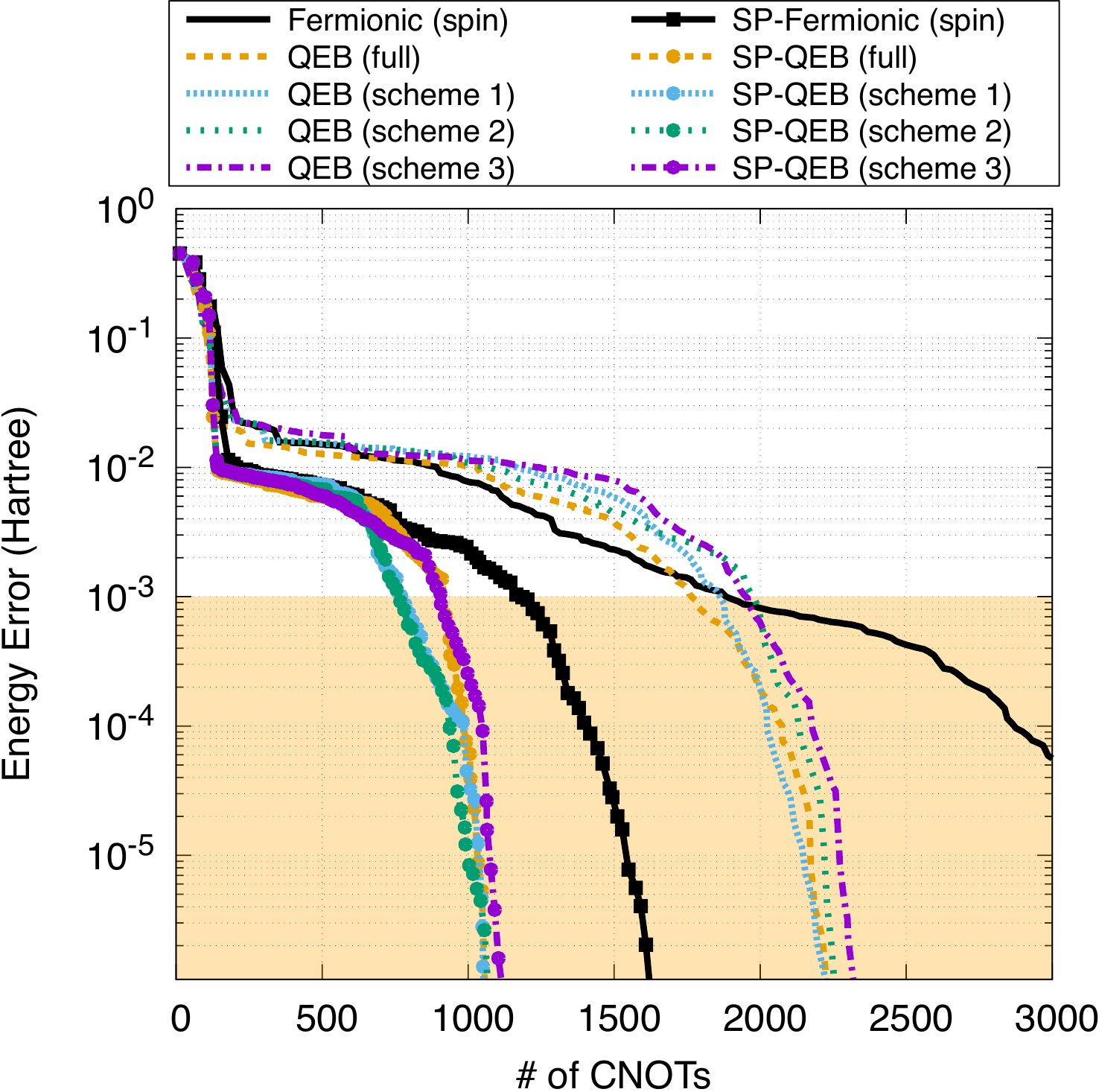}
	\caption{Comparison of the CNOT performance between different operator pools in QEB and SP-QEB.}\label{fig:qeb}
\end{figure}

\begin{figure}
		\includegraphics[width = 25em]{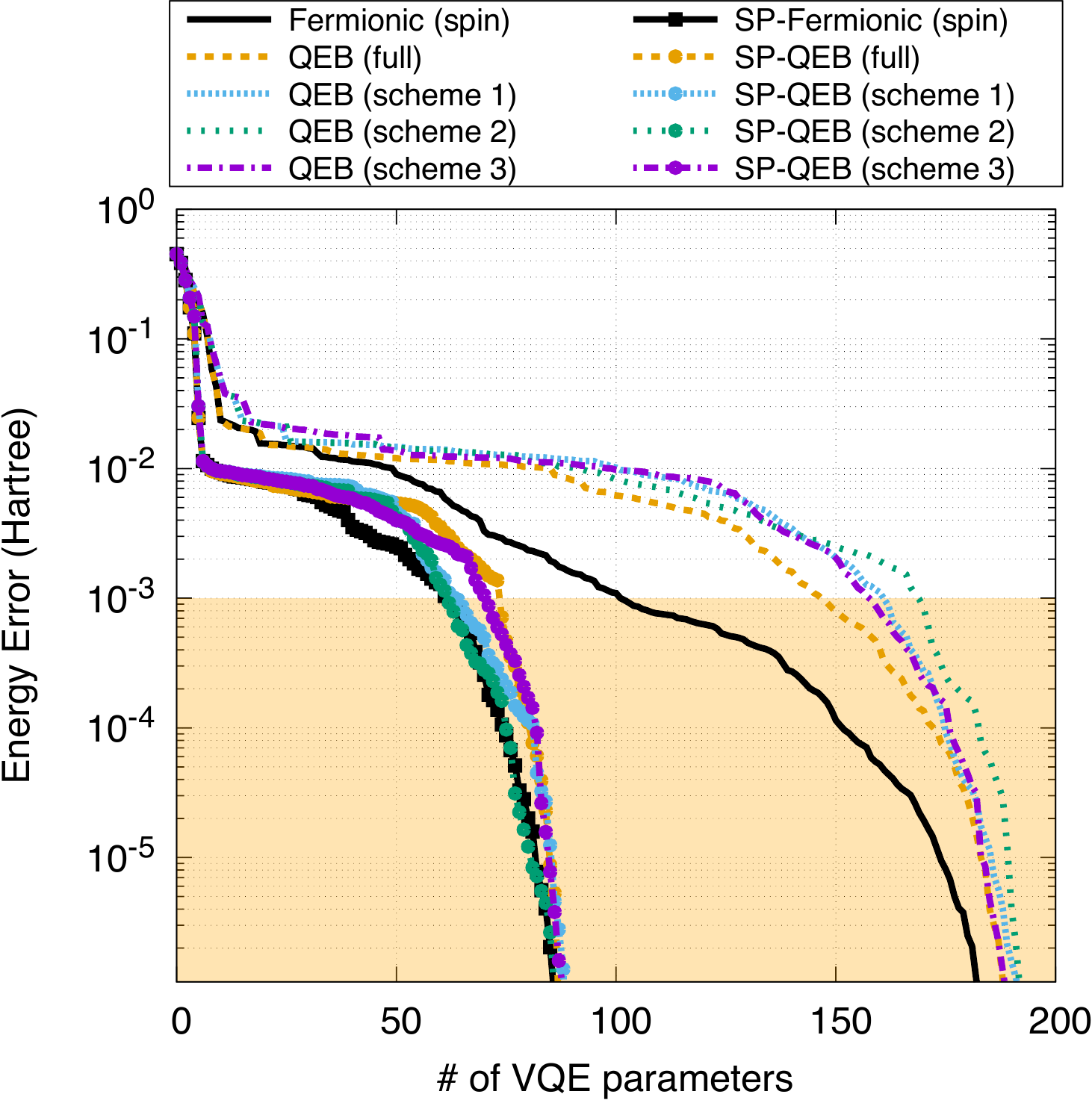}
	\caption{Same as Fig.~\ref{fig:qeb} but as a function of the number of parameters.}\label{fig:qeb_param}
\end{figure}

\subsection{Comparison between different operator pools}\label{sec:pools}
The previous section suggested that, not only is the number of CNOT gates required to obtain the FCI state constantly smaller for QEB-ADAPT than for Fermionic-ADAPT, but also the number of required variational parameters is very similar with spin-projection. We also performed calculations with spin-dependent Fermionic ADAPT using ${\mathscr P}^{\rm spin}$ defined in Eq.~(\ref{eq:SPPool}) and arrived at the same conclusion for broken-symmetry standard ADAPT algorithms. 

The authors of Ref.~[\onlinecite{Tang21}] verified that only $2n-2$ Pauli operators can span the $n$-qubit real Fock space. Although this theorem appears very promising, much more operators are usually required in the ADAPT algorithm because it chooses only one operator at a time; thus, the pool has to be physical in some sense (i.e., it should contain fermionic excitations higher than singles to examine the $k$-particle generalized Brillouin theorem). Nevertheless, the theorem suggests the number of operators in the QEB pool can be further reduced, thereby providing an opportunity to save the computational efforts to choose the next operator. Here, we propose the following three additional qubit-excitation pools and compare their performances:
\begin{itemize}
	\item Scheme 1. $\hat {\tilde \tau}_{q_\alpha}^{p_\alpha}$, $\hat {\tilde \tau}_{r_\beta s_\alpha}^{p_\beta q_\alpha}$ ($\beta$ singles and doubles with the same spins are omitted).
	\item Scheme 2. Same as Scheme 1 but with $p \ge q$.
	\item Scheme 3. Same as Scheme 2 but with $r \ge s$.
\end{itemize}

For the 12-qubit systems, the number of operators in each pool is 855, 555, 400, 190, and 155 for spin-dependent Fermionic, QEB, Scheme 1, Scheme 2, and Scheme 3, respectively. We have plotted the energy convergence of the earlier H$_6$ example against CNOT gates in Fig.~\ref{fig:qeb} and parameter numbers in Fig.~\ref{fig:qeb_param}. From these figures, it can be observed that, at convergence (an energy error of 10$^{-6}$), all QEB schemes perform similarly in terms of both the numbers of CNOT gates and VQE parameters. However, each plot behaves slightly differently. For broken-symmetry ADAPT algorithms (indicated by the lines without points), full QEB provides a slightly better description than Scheme 3 before the convergence. This could be because the entire operator pool in QEB contains more choices than Scheme 3. In particular, the latter does not contain spin-complement operators, which are subsequently chosen in several instances because spin-complement excitations result in similar energy gradients. It is also clear from Fig.~\ref{fig:qeb_param} that spin-dependent Fermionic ADAPT generally exhibits a better energy convergence for a given number of VQE parameters, although it is inefficient in terms of CNOT gate counts. For example, with 100 parameters, the energy obtained by Fermionic ADAPT is one order of magnitude more accurate than those by QEB schemes.

With spin-projection, the performances of different QEB schemes are almost equivalent with a certain CNOT gate count and parameter number. In particular, its accuracy is comparable to SP-Fermionic ADAPT when the number of parameters is fixed, because spin-projection can partially replicate the spin-complement effects. However, SP-Fermionic ADAPT is almost always more accurate than SP-QEB-ADAPT, although slightly. In the particular case of H$_6$, their differences are marginal, and may be rather attributed to the fact that the current approach to determining the next operator is not optimal; operators that are not chosen could identify a lower energy than the chosen one when VQE is performed. However, this behavior is general, as is clearly observed in Fig.~\ref{fig:Error_vs_param}, and can seemingly become more significant for larger systems, as will be demonstrated later. Nevertheless, the improved similarity between the SP-Fermionic and SP-QEB algorithms over standard ADAPT is encouraging for practical applications where the actual operator number (or CNOT gate counts) that can be implemented is assumed to be limited.
\begin{figure}
	\includegraphics[width = 27em]{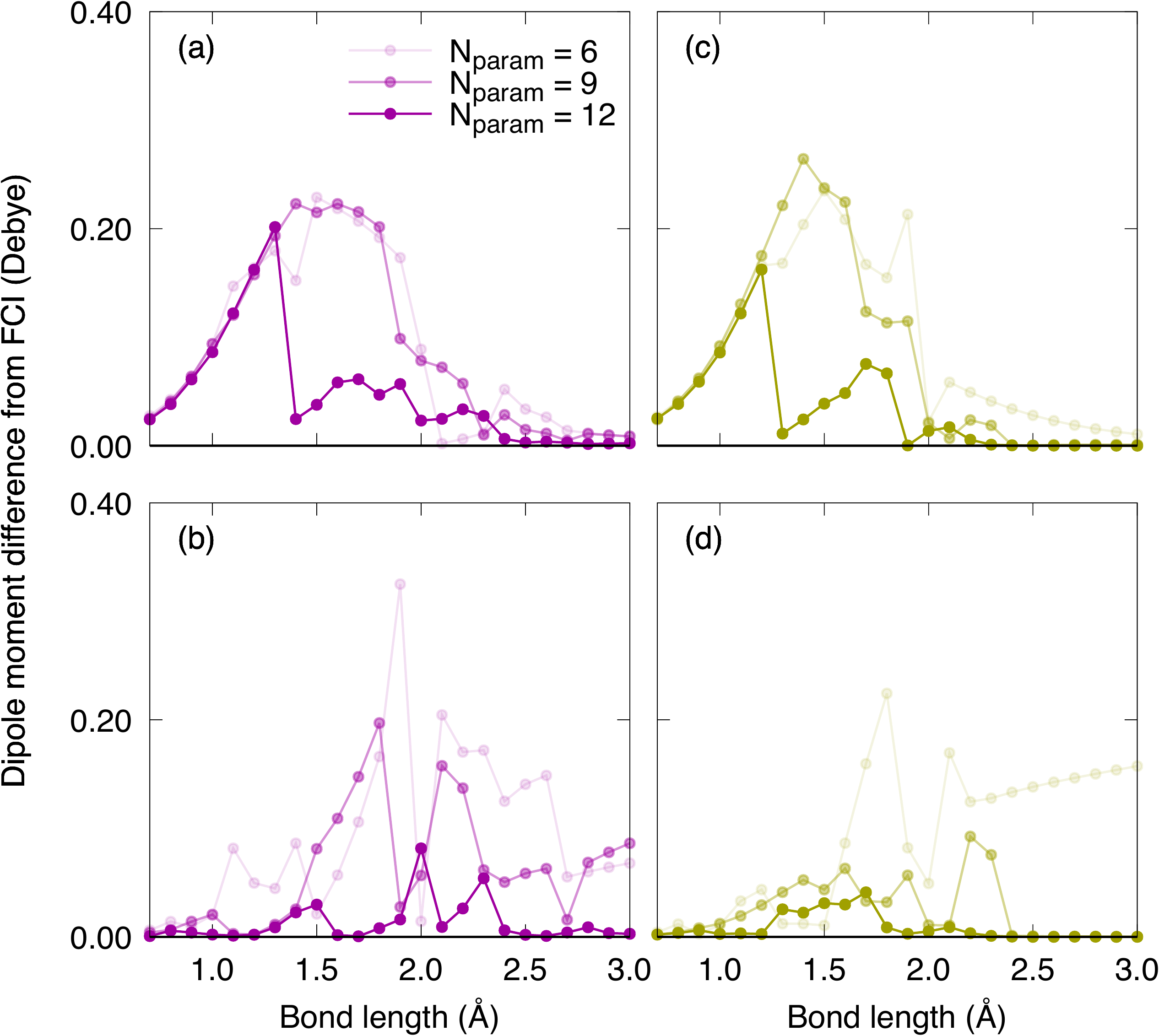}
	\caption{Error from FCI in calculated dipole moment of H$_2$O by altering the number of operators in ADAPT. 
	(a) Spin-dependent Fermionic ADAPT without orbital response, 
	(b) Spin-dependent Fermionic ADAPT with orbital response, 
	(c) SP-Fermionic ADAPT without orbital response, and 
	(d) SP-Fermionic ADAPT with orbital response.}\label{fig:H2O}
	\end{figure}

\begin{figure*}
	\includegraphics[width = 55em]{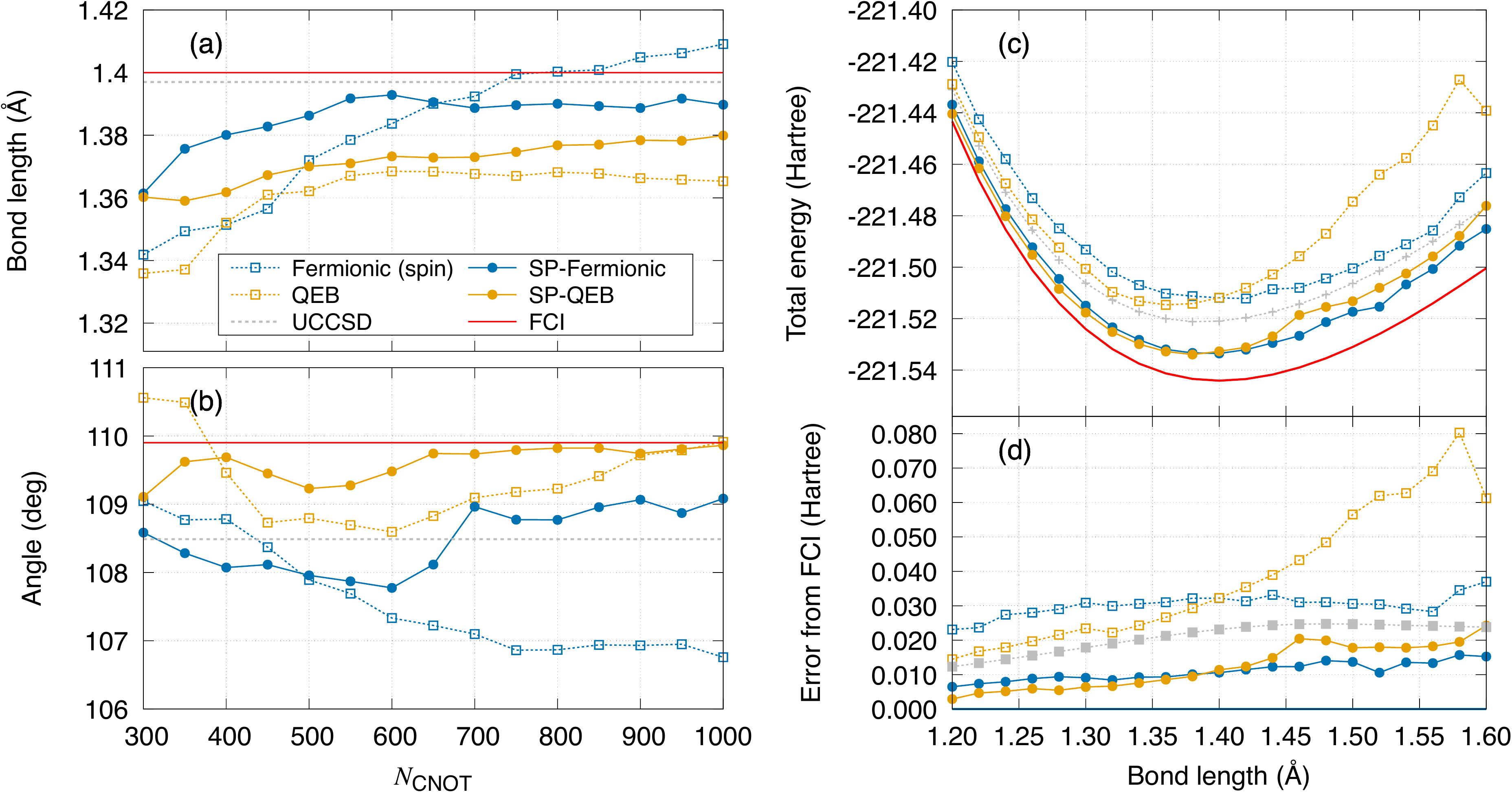}
	\caption{(a) Optimized bond length of O$_3$ as a function of the maximum number of CNOTs used in ADAPT. (b) Same as (a) but for the angle. (c) Potential energy curve of O$_3$ in Hartree, computed with $N_{\rm CNOT}=1000$. (d) Energy error from FCI.}\label{fig:O3}
\end{figure*}
 
\subsection{Molecular properties}\label{sec:property_results}
In this section, we study the molecular properties computed by several ADAPT-based algorithms using the method discussed in Sec.~\ref{sec:properties}. In a practical application of the ADAPT algorithm, we cannot expect to obtain FCI accuracy, but have to make a tradeoff between accuracy and either the circuit depth or the number of measurements. To benchmark their accuracy, we compare the results obtained with the fixed number of VQE parameters, $N_{\rm param}$, or the number of CNOT gates, $N_{\rm CNOT}$. 

\subsubsection{Dipole moment of H$_2$O}
Dipole moment is an important first-order property of a molecule, providing a measure of polarity of  the system. To assess the accuracy of ADAPT in computing this quantity, we have employed the H$_2$O molecule with a fixed angle of $104.5^\circ$ and varied the OH bond length symmetrically. Fig.~\ref{fig:H2O} summarizes the dipole moment error from the FCI value calculated at different ADAPT cycles ($N_{\rm param} = 6, 9,$ and $12$). The results of spin-dependent Fermionic ADAPT without and with the orbital response correction are presented in Figs.~\ref{fig:H2O}(a) and (b), respectively. As can be observed from the figures, the orbital response correction is important for describing the dipole moment correctly at short bond lengths ($\sim$1.5 {\AA}). Note that the correction effect is almost absent in UCCSD because it takes care of single excitations explicitly in VQE by playing the role of orbital relaxation. For the larger bond distance, the orbital response correction sometimes worsens the result; however, this means the ADAPT state is not stationary with respect to orbital rotation and the good dipole moments obtained in the unrelaxed results (Fig.~\ref{fig:H2O}(a)) are fortuitous. Having said that, because we are using the HF canonical orbitals in ADAPT, the orbital Hessian matrix {\bf A} in Eq.~(\ref{eq:R+zA}) can be nearly singular at a certain bond length (around 1.9 {\AA}), where the first-order relaxed density matrix of ADAPT is overcorrected. This singular behavior is circumvented by increasing the number of operators such that the residual $R_{rs}$ is close enough to zero. 

When spin-projection is applied, we essentially observe a similar trend (see Figs.~\ref{fig:H2O}(c,d)). Although spin-symmetry breaking in ADAPT does not affect the estimated dipole moment because both the singlet and contaminated triplet states give almost zero dipole moments at stretched bond lengths, SP-Fermionic ADAPT shows promising results, yielding better convergence with respect to $N_{\rm param}$, especially in a more strongly correlated region (Fig.~\ref{fig:H2O}(d)).

\subsubsection{Geometry and potential curve of O$_3$}
It is difficult to investigate the accuracy in estimating geometrical parameters of strongly correlated systems as we are restricted to small molecules or toy models because of a limited computational budget, and most small molecules are only weakly correlated at equilibrium. In fact, previous studies have considered such simple molecules,\cite{OBrien19, Mitarai20, Delgado21} which do not allow one to properly evaluate the potential of quantum computers. Another issue is that, for adaptive algorithms like ADAPT, the potential energy surface is not smooth, and determining the minimum can be challenging. However, it is interesting to benchmark how effective the geometry optimization in ADAPT methods can be. We think it is also pedagogical to compare the results of spin-dependent Fermionic and QEB ADAPT algorithms in predicting potential energy surfaces when the same quantum resource is available. 

Accordingly, we choose the ozone molecule with the STO-3G basis as our test case, using an active space comprising 9 orbitals and 12 electrons, resulting in an 18-qubit system. To make a direct comparison between the accuracy of each ADAPT method, we have assumed that the number of CNOT gates that one can handle for each calculation is limited (except for UCCSD, where 2534 CNOT gates were required). Namely, all ADAPT results were obtained with less than a certain $N_{\rm CNOT}$ ranging from 300--1000. In all simulations, we tapered qubits to reduce the computational cost but assumed the total number of CNOT gates remain unchanged. 
\begin{table}
\tabcolsep=0.7em
\caption{Non-parallelity error in Hartree.}\label{tb:NPE}
    \begin{tabular}{ccccccc}
        \hline\hline
        Fermionic & QEB & SP-Fermionic& SP-QEB & UCCSD \\
         \hline
        0.014 & 0.066 & 0.009 & 0.021 & 0.012\\
        \hline\hline
        \end{tabular}
\end{table}

Figs.~\ref{fig:O3}(a) and (b) depict the optimized bond length ({\AA}) and angle ($^\circ$), where the FCI result is depicted by the red line (R$_{\rm O-O} = 1.400$ {\AA} and $\angle {\rm OOO} = 109.9^\circ$). Because O$_3$ is a two-determinant system, UCCSD is reasonably accurate, as indicated by the dotted gray line. Thus, symmetry-breaking in ADAPT is not so significant for both the Fermionic and QEB states. Our calculations indicate that $\langle \hat S^2\rangle$ in ADAPT is approximately $0.1$ at most, at the equilibrium geometry estimated by FCI. From the figures, the bond length is more accurately predicted by Fermionic ADAPT than by QEB ADAPT, with or without spin-projection, while the results for the angle exhibit the opposite trend. Without spin-projection, the two methods yielded significantly different results for geometry optimization, especially at a larger $N_{\rm CNOT}$. Increasing $N_{\rm CNOT}$ does not always lead to more accurate results, at least up to $N_{\rm CNOT} = 1000$. However, by applying spin-projection, it appears the geometries predicted by the SP-Fermionic and SP-QEB algorithms are improved with $N_{\rm CNOT}$. They are also similar to each other, implying the two methods also result in similar states. 

To further investigate these results, we have plotted in Fig.~\ref{fig:O3}(c) the potential energy curve of symmetric bond dissociation of O$_3$ with a fixed angle of 109.9$^\circ$, using $N_{\rm CNOT} = 1000$. In Fig.~\ref{fig:O3}(d), we have also presented the energy error from FCI in Hartree. In all the methods, to a certain degree, the energy is more accurate at a shorter (weakly correlated) distance and becomes worse as the bond is stretched (strongly correlated). While this result is quite reasonable, we found that the change in QEB energy is rather  significant compared to that in the Fermionic methods, especially without spin-projection. When comparing the Fermionic and QEB results, their energy accuracy is inverted at 1.40 {\AA}. We took the non-parallelity error (NPE) of the potential energy curve, which is defined as the difference between the maximum and minimum errors from FCI throughout the potential curve, and tabulated in Table \ref{tb:NPE}. As this table shows, Fermionic ADAPT provides a more parallel curve to that of FCI than QEB ADAPT. This explains why the latter is less accurate than the former in predicting the bond length, as illustrated in Fig.~\ref{fig:O3}(a).

\begin{figure}
		\includegraphics[width = 25em]{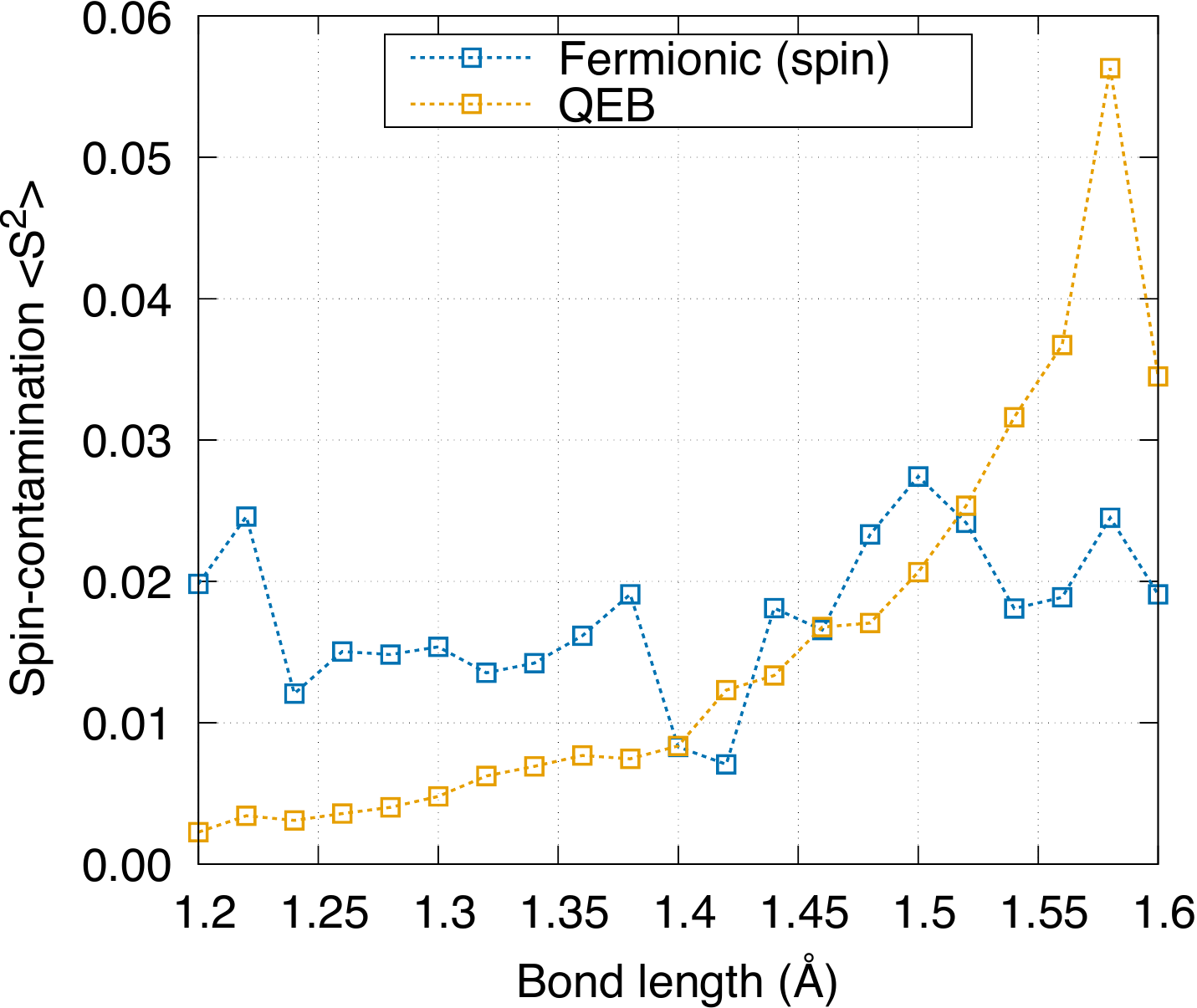}
	\caption{Spin-contamination of Fermionic and QEB ADAPT in O$_3$.}\label{fig:O3_S2}
\end{figure}

 At this point, we are uncertain whether this failure of QEB can be attributed to its fundamental deficiency or the possibly inappropriate choice of operators in the ADAPT algorithm. Because the deviation between Fermionic and QEB ADAPT results are significantly mitigated by spin-projection (Figs.~\ref{fig:O3}(c,d)), we suspect symmetry breaking to be behind the different behaviors. Therefore, in Fig.~\ref{fig:O3_S2}, we have summarized the $\langle \hat S^2\rangle$ values obtained from the calculations in Fig.~\ref{fig:O3}(c). In fact, we find that, although Fermionic ADAPT retains the same degree of symmetry breaking throughout all bond distances, QEB is more spin-contaminated when the molecule is stretched. As we have seen several times, symmetry breaking can slow down the convergence in ADAPT; therefore, the degree of spin-contamination is directly related to the energy accuracy with a fixed $N_{\rm CNOT}$. Hence, spin-projection plays an important role in equalizing the two schemes to some extent.

\section{Conclusions}\label{sec:conclusions}
One of the unsolved problems in quantum chemistry is a consensus way of treating strong correlations. Quantum computing is expected to provide a solution to this problem because it can, in principle, handle entangled quantum states directly mapped onto qubits. The recently proposed ADAPT algorithm has paved the way to construct an efficient quantum circuit adaptively; however, its convergence is slow for strongly correlated systems because of a significant symmetry breaking. 

To address this issue, we introduced spin-projection to ADAPT. We demonstrated that spin-projection can be quite effective when combined with ADAPT, offering shallower circuits with lesser CNOT gates and VQE parameters to achieve the same accuracy at the cost of increased measurements. This is especially the case if a gate-efficient circuit is adopted for fermionic and qubit excitations, whose performances are often comparable. Our calculations also indicated that the pool of QEB ADAPT can be reduced by discarding certain classes of spin-dependent qubit excitations. However, we concluded that it is not worthwhile to apply symmetry-projection to qubit-ADAPT because it would require the restoration of lost $\hat S_z$ (and number) symmetry, which considerably increases the number of measurements in evaluating the energy. 

We have also derived the first-order energy derivative in the presence of spin-projection, which enabled the calculation of dipole moment and geometry optimization with SP-ADAPT. It was demonstrated that the orbital response correction is important both for ADAPT and SP-ADAPT in the calculation of dipole moment, because they are far from stationary with respect to orbital changes, unlike UCCSD, which is less sensitive to orbital rotation because of the presence of explicit single excitations. Furthermore, the method was applied to the geometry optimization of O$_3$ to quantify the capabilities of Fermionic and QEB schemes. We found that QEB ADAPT is less stable in achieving a constant accuracy throughout the potential energy surface and thus is less predictable than Fermionic ADAPT in terms of optimized geometry. The deviation between the two schemes was caused by a larger spin-contamination in QEB ADAPT and could be largely mitigated by performing spin-projection. However, the reason behind this unfavorable behavior of QEB ADAPT remains unclear, and further studies will be required to elucidate its cause.

\section*{Acknowledgements}
This work was supported by JST, PRESTO Grant Number JPMJPR2016, Japan and also by JSPS KAKENHI, Grant Numbers JP20K15231 and JP18H03900. We are grateful for the computational resources provided by ECCSE, Kobe University.
\appendix

\section{First-order energy derivative of VQE}\label{appendix}
\subsection{General strategy}
Here, we discuss the most general formulation for the first-order derivative of energy to obtain the relaxed density matrix in variational methods. Although the derivation is known, we believe it is still beneficial to re-derive and summarize the equations for non-experts. As shown below, we adopt spin molecular orbitals instead of spatial molecular orbitals to ensure that our derivation is as general as possible. With the canonical HF orbitals, the Fock matrix 
\begin{align}
	F_{pq} &= h_{pq} + \sum_{rs}\langle pr||qs\rangle D^{\rm HF}_{rs}
\end{align} 	
is diagonal, where we define the idempotent HF density matrix
\begin{align}
	D^{\rm HF}_{rs}&=\langle \Phi^{\rm HF}| a^\dag_r a_s | \Phi^{\rm HF}\rangle = \begin{cases}
		1 & (r =s \in occ)\\
		0 & (r \ne s)
	\end{cases}
\end{align} 
Therefore, it is convenient to use off-diagonal elements to constrain the Lagrangian in Eq.~\ref{eq:L}. Because canonical HF orbitals have occupied and virtual spaces, we decompose the Fock elements in ${\cal L}$ for convenience as
\begin{align}
	{\cal L}[{\bm\theta},{\bm\kappa}, {\bf z}, x] = E[{\bm\theta},{\bm\kappa},x] + \sum_{a}^{vir}\sum_{i}^{occ} z_{ai} F_{ai} + \sum_{i>j}^{occ} z_{ij} F_{ij} + \sum_{a>b}^{vir} z_{ab} F_{ab},
\end{align}
where $i,j,k,l$ denote occupied orbitals and $a,b,c,d$ denote virtual orbitals. 
The requirements for the orbital derivative are
\begin{subequations}
\begin{align}
	\frac{\partial \cal L}{\partial \kappa_{ck}} = 0 \label{eq:dLdKck}\\
	\frac{\partial \cal L}{\partial \kappa_{kl}} = 0 \label{eq:dLdKkl}\\
	\frac{\partial \cal L}{\partial \kappa_{cd}} = 0	 \label{eq:dLdKcd}
\end{align}
\end{subequations}
and therefore, we compute the Fock derivative $A_{pq,rs} = \partial F_{pq}/\partial \kappa_{rs}$ in each orbital sector. The results are
\begin{subequations}
\begin{align}
A_{ai,ck} &= (F_{aa} - F_{ii}) \delta_{ac}\delta_{ik} + \langle ak||ic\rangle + \langle ac||ik\rangle \label{eq:Aaick}\\
A_{ij, ck} &= \langle ic||jk\rangle + \langle ik || jc\rangle \label{eq:Aijck}\\
A_{ab,ck} & = \langle ac||bk\rangle + \langle ak||bc\rangle\label{eq:Aabck}\\
A_{ij,kl} &= (F_{ii} - F_{jj}) (\delta_{ik} \delta_{jl} - \delta_{il}\delta_{jk}) \label{eq:Aijkl}\\
A_{ab, cd} &= (F_{aa} - F_{bb}) (\delta_{ac}\delta_{bd} - \delta_{ad}\delta_{bc}) \label{eq:Aabcd}
\end{align}\label{eq:Amat}
\end{subequations}
and all other terms are strictly zero. 

We first solve Eq.~(\ref{eq:dLdKkl}). Because $i>j$ and $k>l$, $\delta_{il}$ and $\delta_{jk}$ are always zero in Eq.~(\ref{eq:Aijck}); hence,
\begin{align}
	\frac{\partial \cal L}{\partial \kappa_{kl}} &= \frac{\partial E}{\partial \kappa_{kl}} + \sum_{i>j}^{occ} z_{ij} A_{ij,kl} \nonumber\\
	&= R_{kl} + \sum_{i>j}^{occ} z_{ij} (F_{ii} - F_{jj}) \delta_{ik}\delta_{jl} =0,
\end{align}
which leads to 
\begin{align}
	z_{ij} = -\frac{R_{ij}}{F_{ii} - F_{jj}}.
\end{align}
Similarly, solving Eq.~(\ref{eq:dLdKcd}) results in 
\begin{align}
	z_{ab} = -\frac{R_{ab}}{F_{aa} - F_{bb}}.
\end{align}
It should be mentioned that if orbitals are (nearly-)degenerate, the denominator can become numerically zero; however, in such a case, we expect the numerator to be approximately zero, meaning the rotation is redundant and can be discarded. We have not faced such a numerically challenging situation.

Now, from Eq.~(\ref{eq:dLdKck}), we find
\begin{align}
	\frac{\partial \cal L}{\partial \kappa_{ck}} &= R_{ck} + \sum_{a}^{vir}\sum_i^{occ} z_{ai}A_{ai,ck} + \sum_{i>j} z_{ij} A_{ij,ck} +\sum_{a>b} z_{ab} A_{ab,ck}\nonumber\\
	&= R_{ck} + R^{occ}_{ck} + R^{vir}_{ck} + \sum_{a}^{vir}\sum_i^{occ} z_{ai}A_{ai,ck} =0
\end{align}
where 
\begin{align}
	R^{occ}_{ck} &= -\sum_{i>j}\frac{R_{ij}}{F_{ii}- F_{jj}}\Bigl( \langle ic||jk\rangle + \langle ik ||jc\rangle\Bigr) 	\\
	R^{vir}_{ck} &= -\sum_{a>b}\frac{R_{ab}}{F_{aa} - F_{bb}}\Bigl( \langle ac||bk\rangle + \langle ak ||bc\rangle\Bigr).
\end{align}
Therefore, we solve
\begin{align}
	\sum_{a}^{vir}\sum_i^{occ} z_{ai}A_{ai,ck} + \tilde R_{ck} = 0
\end{align}
with 
\begin{align}
	\tilde R_{ck} \equiv R_{ck} + R^{occ}_{ck} + R^{vir}_{ck}
\end{align}
to determine $z_{ai}$.

Finally, the relaxed density matrices $D^{\rm relax}_{pq}$ and $D^{\rm relax}_{pq,rs}$ are obtained by comparing the one-body and two-body terms in ${\cal L}$,
\begin{align}
	{\cal L} &= \sum_{pq} h_{pq}\langle \psi|a_p^\dag a_q| \psi\rangle + \frac{1}{4}\sum_{pqrs}\langle pq||rs\rangle \langle \psi|a_p^\dag a_q^\dag a_s a_r| \psi\rangle \nonumber\\
	&+ \sum_{p>q} z_{pq} \left( h_{pq} + \langle pr||qs\rangle \langle \Phi^{\rm HF}| a^\dag_r a_s | \Phi^{\rm HF}\rangle\right)\nonumber\\
	&= \sum_{pq} h_{pq} D^{\rm relax}_{pq} + \frac{1}{4}\sum_{pqrs}\langle pq||rs\rangle D^{\rm relax}_{pq,rs}
\end{align}
where $|\psi\rangle$ represents the VQE wave function and
\begin{align}
	D^{\rm relax}_{pq} &= D_{pq} + \frac{1}{2}z_{pq}\\
	D^{\rm relax}_{pq,rs}&= D_{pq,rs} + \frac{1}{2} \Big( z_{pr} D^{\rm HF}_{qs} - z_{qr} D^{\rm HF}_{ps} \nonumber\\
	&\;\;\;\;\;\;- z_{ps} D^{\rm HF}_{qr} + z_{qs} D^{\rm HF}_{pr} \Big)\\
	D_{pq} & =\langle \psi|a_p^\dag a_q| \psi\rangle\\
	D_{pq,rs} &= \langle \psi|a_p^\dag a_q^\dag a_s a_r| \psi\rangle
	\end{align}
Note that $D^{\rm relax}_{pq,rs}$ is explicitly anti-symmetrized such that $D^{\rm relax}_{pq,rs} = - D^{\rm relax}_{qp,rs}$, and so on, for convenience. 

\subsection{Frozen-core approximation}
When the frozen-core approximation is exercised, the VQE energy derivative with respect to the mixing between the frozen-core $I$ and active $p$ orbitals (i.e., those mapped to qubit), $R_{pI}$, cannot be directly evaluated by qubit measurements. Nevertheless, we can compute these values using the density matrices within the active space as follows: 
\begin{align}
	R_{pI} &= 2\Big( \tilde h_{Ip} - \sum_{q}^{ACT} D_{qp}\tilde h_{qI} + \sum_{qr}^{ACT} \langle Iq||pr\rangle D_{qr} \nonumber\\
	&- \frac{1}{2}\sum_{qrs}^{ACT} \langle qr||sI\rangle D_{qr,sp} \Big)
\end{align}
where we have defined the core Hamiltonian
\begin{align}
	\tilde h_{PQ} = h_{PQ} + \sum_I^{Frozen-core}\langle PI||QI\rangle
\end{align}
with general spin-orbitals $P,Q$.

The Fock derivatives and multipliers with frozen-core indices, such as $A_{aI, ck}$ and $z_{cI}$, are computed with the same equations as above, simply by expanding $i,j,k,l$ to include the frozen-core orbitals.

For the frozen-virtual approximation, we must consider the contribution from the frozen-virtual orbitals $A$:
\begin{align}
	R_{Ap} &= 2\left(\sum_{q}^{ACT} D_{pA}\tilde h_{qA} + \frac{1}{2}\sum_{qrs}^{ACT} \langle qr||sA\rangle D_{qr,sp} \right)\\
	R_{AI} &= 2 \left(\tilde h_{IA} + \sum_{pq}^{ACT} \langle pI|| qA\rangle D_{qp}\right)
\end{align}
\subsection{Nuclear gradients}
In several cases, HF orbitals are expressed as a linear combination of atomic orbitals (AO) that are functions of atomic coordinates. If AOs depend on the perturbation $x$ (nuclear displacement), the electronic energy derivative includes a contribution from the AO overlap derivative $S^{(x)} = \partial \langle \phi_\mu|\phi_\nu\rangle/\partial x$. The general expression is 
\begin{align}
	\frac{dE}{dx} &= \sum_{\mu\nu} h^{(x)}_{\mu\nu}D^{\rm relax}_{\mu\nu} + \frac{1}{4} \sum_{\mu\nu\lambda\sigma}\langle \mu\nu||\lambda\sigma\rangle^{(x)} D^{\rm relax}_{\mu\nu,\lambda\sigma} 
	\nonumber\\&+\sum_{\mu\nu} S^{(x)}_{\mu\nu}W_{\mu\nu} 
\end{align}
where the superscript $(x)$ represents the (explicit) partial derivative, and $W_{\mu\nu}$ denotes the so-called energy weighted density matrix,
\begin{align}
	W_{\mu\nu} = -\left(\sum_{\lambda}h_{\mu\lambda} D^{\rm relax}_{\nu\lambda} + \frac{1}{2}\sum_{\lambda}\langle \mu \kappa ||\lambda\sigma\rangle D^{\rm relax}_{\nu \kappa, \lambda \sigma} \right).
\end{align}

\subsection{Spin-projection}
When the spin-projection operator $\hat P$ is present, the same procedure is followed, but with Eq.~(\ref{eq:RmSP}) instead of Eq.~(\ref{eq:Rm}) for $R_{pq}$. As a caveat, for spin-dependent properties such as hyperfine coupling constants, the relaxed density matrices need to be slightly modified. Our SP-VQE energy is expressed as 
\begin{align}
	E & = \frac{1}{\langle \psi| \hat P |\psi\rangle} \Bigl(\sum_{pq} h_{pq}\langle \psi|a_p^\dag a_q\hat P| \psi\rangle \nonumber\\&+ \frac{1}{4}\sum_{pqrs}\langle pq||rs\rangle \langle \psi|a_p^\dag a_q^\dag a_s a_r \hat P| \psi\rangle\Bigr)
\end{align}
Here, notice that $\langle \psi|a_p^\dag a_q\hat P| \psi\rangle/\langle \psi| \hat P |\psi\rangle$ does {\it not} correspond to the genuine density matrix of SP-VQE, that is, $\langle \psi|\hat P^\dag a_p^\dag a_q\hat P| \psi\rangle \ne \langle \psi|a_p^\dag a_q\hat P| \psi\rangle$. In fact, the ``half-projected'' density matrix and its response correction are not spin-adapted, and therefore, lead to an incorrect spin density. To avoid this issue, the Wigner-Eckart theorem can be adopted, which is expressed as
\begin{widetext}
\begin{align}
\hat P^s_{m,m} (a^\dag_{p\alpha}a_{q\alpha} + a_{p\beta}^\dag a_{q\beta}) \hat P^s_{m,m}
&=	\langle s\:m\:0\:0|s\:m\rangle^2 (a_{p\alpha}^\dag a_{q\alpha} + a_{p\beta}^\dag a_{q\beta}) \hat P_{m,m}^s\\
\hat P^s_{m,m} (a^\dag_{p\alpha}a_{q\alpha} + a_{p\beta}^\dag a_{q\beta}) \hat P^s_{m,m}
&= \langle s\: m\: 1\: 0| s\: m\rangle \Big(\langle s\: m\: 1\: 0| s\: m\rangle (a_{p\alpha}^\dag a_{q\alpha}-a_{p\beta}^\dag a_{q\beta})\hat P_{m,m}^s \nonumber\\
	&+ \langle s\: m\: 1\: -1| s\: m\rangle a_{p\alpha}^\dag a_{q\beta} \hat P_{m-1,m}^s - \langle s\: m\: 1\: 1| s\: m\rangle (a_{p\beta}^\dag a_{q\alpha}) \hat P_{m+1,m}^s\Bigr)
\end{align}
\end{widetext}
where $\langle s_1\:m_1\:s_2\: m_2|S\:M\rangle$ represents the Clebsch-Gordan coefficient and $\hat P^s_{m,k} = |s;m\rangle \langle s;k|$ denotes the transfer operator, which is a general form of the spin-projection operator $\hat P \equiv \hat P^s_{m,m}$.

\bibliographystyle{apsrev4-1}

%
\end{document}